\documentclass[amsmath,showpacs,nofootinbib,12pt]{revtex4-2}
\usepackage{graphicx}
\usepackage{dcolumn}
\usepackage{bm}
\usepackage{color} 
\usepackage{slashed}
\DeclareUnicodeCharacter{2212}{-}
\begin{document}
\newcommand{\hs}{\hspace*{0.5cm}}
\newcommand{\vs}{\vspace*{0.5cm}}
\newcommand{\be}{\begin{equation}}
\newcommand{\ee}{\end{equation}}
\newcommand{\bea}{\begin{eqnarray}}
\newcommand{\eea}{\end{eqnarray}}
\newcommand{\ben}{\begin{enumerate}}
\newcommand{\een}{\end{enumerate}}
\newcommand{\bde}{\begin{widetext}}
\newcommand{\ede}{\end{widetext}}
\newcommand{\nn}{\nonumber}
\newcommand{\crn}{\nonumber \\}
\newcommand{\Tr}{\mathrm{Tr}}
\newcommand{\non}{\nonumber}
\newcommand{\noi}{\noindent}
\newcommand{\al}{\alpha}
\newcommand{\la}{\lambda}
\newcommand{\bet}{\beta}
\newcommand{\ga}{\gamma}
\newcommand{\va}{\varphi}
\newcommand{\om}{\omega}
\newcommand{\pa}{\partial}
\newcommand{\+}{\dagger}
\newcommand{\fr}{\frac}
\newcommand{\bc}{\begin{center}}
\newcommand{\ec}{\end{center}}
\newcommand{\Ga}{\Gamma}
\newcommand{\de}{\delta}
\newcommand{\De}{\Delta}
\newcommand{\ep}{\epsilon}
\newcommand{\varep}{\varepsilon}
\newcommand{\ka}{\kappa}
\newcommand{\La}{\Lambda}
\newcommand{\si}{\sigma}
\newcommand{\Si}{\Sigma}
\newcommand{\ta}{\tau}
\newcommand{\up}{\upsilon}
\newcommand{\Up}{\Upsilon}
\newcommand{\ze}{\zeta}
\newcommand{\ps}{\psi}
\newcommand{\Ps}{\Psi}
\newcommand{\ph}{\phi}
\newcommand{\vph}{\varphi}
\newcommand{\Ph}{\Phi}
\newcommand{\Om}{\Omega}
\newcommand{\AdrHEPC}{Phenikaa Institute for Advanced Study and Faculty of Basic Science, Phenikaa University, Yen Nghia, Ha Dong, Hanoi 100000, Vietnam}

\title{Flavor-dependent $U(1)$ extension inspired by lepton, baryon and color numbers} 

\author{Duong Van Loi}
\email{loi.duongvan@phenikaa-uni.edu.vn (corresponding author)}
\author{Phung Van Dong} 
\email{dong.phungvan@phenikaa-uni.edu.vn}

\affiliation{\AdrHEPC} 
\date{\today}

\begin{abstract}
There is no reason why the gauge symmetry extension is family universal as in the standard model and the most well-motivated models, e.g. left-right symmetry and grand unification. Hence, we propose a simplest extension of the standard model---a flavor-dependent $U(1)$ gauge symmetry---and find the new physics insight. For this aim, the $U(1)$ charge, called $X$, is expressed as $X=x B+y L$ in which $x$ and $y$ are free parameters as functions of flavor index, e.g. for a flavor $i$ they take $x_i$ and $y_i$ respectively, where $B$ and $L$ denote normal baryon and lepton numbers. Imposing a relation involved by the color number $3$, i.e. $-x_{1,2,\cdots,n}=x_{n+1,n+2,\cdots,n+m}=3y_{1,2,\cdots,n+m}\equiv 3z$, for arbitrarily nonzero $z$, we achieve a novel $U(1)$ theory with implied $X$-charge. This theory not only explains the origin of the number of observed fermion families but also offers a possible solution for both neutrino mass and dark matter, which differs from $B-L$ extension. Two typical models based on this idea are examined, yielding interesting results for flavor-changing neutral currents and particle colliders, besides those of neutrino mass and dark matter. 
\end{abstract}

\maketitle

\section{Introduction}
Although the standard model (SM) has been remarkably successful in describing the fundamental particles and interactions, it is known to be an incomplete theory and requires extension. Problems considered as limitations/shortcomings of the SM include the existence of dark matter which makes up most mass of galaxies and galaxy clusters in our universe~\cite{Planck:2018vyg}, neutrino oscillations which require non-zero neutrino masses and flavor mixing \cite {RevModPhys.88.030501, RevModPhys.88.030502}, and specially why there are just three fermion families observed in the nature \cite{ParticleDataGroup:2022pth}.

One of the most interesting extensions of the SM is an extra gauge symmetry $U(1)$ with associated gauge boson $Z'$. Such extension is well motivated since it not only occurs in grand unification and string theories but also supplies a new gauge dynamics which accounts the SM issues and experimental deviations \cite{Langacker:2008yv}, for instance, solution for dark matter \cite{Foot:2014uba,Okada:2016tci,Okada:2018tgy,Dutta:2019fxn,Choi:2020nan,Jho:2020sku,Okada:2020evk,VanDong:2020bkg,VanDong:2020cjf,VanLoi:2020kdk,Okada:2022cby}, neutrino mass generation \cite{Bertuzzo:2018ftf,Das:2018tbd,Asai:2018ocx,Das:2019pua,Ghosh:2021khk}, muon anomalous magnetic moment \cite{Ma:2001md,Baek:2001kca,Baek:2008nz}, asymmetric matter production \cite{VanDong:2020nwb}, and fermion family number \cite{VanDong:2022cin,VanLoi:2023bcp}. It is stressed that this type of extension can be distinguished by $Z'$ mass, $Z'$-fermion couplings, extended Higgs sector, possible couplings to a hidden sector, and kinetic mixing effect, etc. 

In this work, we propose a flavor-dependent $U(1)$ extension of the SM, which has rarely been considered in the literature \cite{Langacker:2000ju}. The $U(1)$ charge, labeled as $X$, has the form, $X=xB+yL$, where free parameters $x$ and $y$ take values $x_i$ and $y_i$ for flavor $i$, respectively, while $B\ (L)$ denotes baryon (lepton) number, as usual. For brevity, we restrict the parameter space by assuming that only $x_i$ depends on flavor $i$, whereas $y_i$ does not, i.e. $y_i=z$ for all $N_f$ families, where $z$ is arbitrarily nonzero. Additionally, the $x_i$ parameters are separated into two types of opposite signs, which relate to the parameter $z$ via the color number $3$, namely, $x_i=-3z$ for the first $n$ families, while $x_i=3z$ for the remaining $m$ families, where $n+m=N_f$. Intriguingly, we obtain a novel $U(1)$ theory that not only requires the presence of right-handed neutrinos due to the anomaly cancellation (a framework for generating nonzero neutrino masses) but also leads to the number of fermion families to be a multiple of the color number, $3$, i.e., $N_f=3(n-m)$. It is noted that QCD asymptotic freedom requires $N_f\leq 8$, which leads two solutions $N_f=3$ and $6$. Imposing the first solution $N_f=3$ as observed, we obtain $n=2$ and $m=1$. Besides supplying a potential explanation to existence of three fermion families, this theory reveals possible answers for neutrino mass and dark matter. We will consider two realistic models recognizing this feature. The first model, called conventional $U(1)_X$ model, predicts tiny neutrino masses via seesaw mechanism as well as scenario of a single dark matter. The second model, called alternative $U(1)_X$ model, provides a scenario of two-component dark matter, by contrast, while nonzero neutrino masses are induced by scotogenic mechanism. In addition, we will discuss interesting implications of each model for flavor-changing neutral currents (FCNCs) and particle colliders in detail. 

The rest of this work is organized as follows: In Sec. \ref{sec2}, we consider anomaly cancellation conditions for the new gauge symmetry and arguments that lead to two realistic models. The implications of each model for neutrino mass and dark matter are presented in Secs. \ref{sec3} and \ref{sec4}, respectively. In Sec. \ref{sec5}, we obtain constraints for each model from flavor-changing neutral currents and particle colliders. Dark matter phenomenology in each model is presented in Sec. \ref{sec6}. Lastly, we summarize the results and make conclusion in Sec. \ref{sec7}.

\section{\label{sec2}$U(1)_X$ anomaly cancellation}

As stated, we extend the SM by adding an extra $U(1)_X$ gauge symmetry, such as
\be SU(3)_C\otimes SU(2)_L\otimes U(1)_Y\otimes U(1)_X.\label{gaugesymmetry} \ee 
The new charge, labeled as $X$, combines lepton and baryon numbers,  
\be X=x B+y L,\ee and depends on family. That said, when $X$ acts on a flavor $i$, it becomes $X_i= x_i B +y_i L$, where $x_i$ and $y_i$ are functions of $i$ for $i=1,2,\cdots,N_f$, while $B\ (L)$ denotes total baryon (lepton) number. The standard model fermions transform under the gauge symmetry (\ref{gaugesymmetry}) as
\bea l_{iL} &=&(\nu_{iL},e_{iL})^T\sim({\bf 1}, {\bf 2}, -1/2, y_i),\label{contenfer1}\\
e_{iR} &\sim& ({\bf 1}, {\bf 1}, -1, y_i),\\
q_{iL} &=&(u_{iL},d_{iL})^T\sim({\bf 3}, {\bf 2}, 1/6, x_i/3),\\
u_{iR} &\sim& ({\bf 3}, {\bf 1}, 2/3, x_i/3),\hs d_{iR}\sim ({\bf 3}, {\bf 1}, -1/3, x_i/3).\label{contenfer2}
 \eea

First, the anomaly $[SU(2)_L]^2U(1)_X\sim\sum_{\mathrm{doublets}}X_{f_L}$ vanishes if $\sum_i(x_i+y_i)=0$. This equation contains $2N_f$ unknowns because of $i=1,2,\cdots,N_f$ and has an infinite number of solutions. Different solutions can be derived by imposing some suitable assumptions, which would lead to distinct kinds of new physics. For example, the $U(1)_{B-L}$ model with $x_i=-y_i=1$ universally for every $i$ makes the anomaly vanishing within each family similar to the SM, and in this case, both the $U(1)_{B-L}$ model and the SM cannot explain the existence of just three fermion families. As the perspective of this work, we assume that only $x_i$ parameters depend on family, whereas $y_i$ parameters do not, i.e. $y_i=z$ for every $i$, where $z$ is arbitrarily nonzero. Additionally, the $x_i$ parameters relate to the $y_i$ parameters via the color number $3$, namely, $x_{1,2,\cdots,n}=-3z$ and $x_{n+1,n+2,\cdots,n+m}=3z$, where $n+m=N_f$. Hence, the anomaly cancellation becomes $3[n(-z)+mz]+N_fz=0$, or equivalently 
\be N_f=3(n-m).\label{Nf} \ee
Since $n$ and $m$ are integer numbers, Eq. (\ref{Nf}) implies the number of fermion families to be a multiple of the color number, $3$. On the other hand, the QCD asymptotic freedom condition requires the family number to be less than or equal to $8$, thus $N_f=3$ and 6. The realistic solution is $N_f=3$ as coinciding with experiment \cite{ParticleDataGroup:2022pth}, which implies $n=2$ and $m=1$. For convenience in reading, we use two kinds of fermion family indices, $a,b=1,2,3$ according to $N_f=3$ and $\al,\bet=1,2$ according to $n=2$, hereafter.

With the fermion content as in Eqs. (\ref{contenfer1})-(\ref{contenfer2}), the $[\mathrm{Gravity}]^2U(1)_X$ and $[U(1)_X]^3$ anomalies are not canceled, which are given by 
\bea [\mathrm{Gravity}]^2U(1)_X &\sim& \sum_{\mathrm{fermions}}(X_{f_L}-X_{f_R})=3z,\\
\left[U(1)_X\right]^3 &\sim&\sum_{\mathrm{fermions}}(X^3_{f_L}-X^3_{f_R})=3z^3.\eea
Similar to the $U(1)_{B-L}$ extension, to cancel these anomalies, right-handed neutrinos that transform nontrivially under the gauge symmetry (\ref{gaugesymmetry}), i.e. $\nu_{iR}\sim({\bf 1},{\bf 1},0,X_{\nu_{iR}})$ for $X_{\nu_{iR}}\neq 0$, are necessarily introduced as fundamental fields, where $i=1,2,\cdots,N_R$ with $N_R$ is the number of right-handed neutrinos added. Additionally, their $X_{\nu_{iR}}$ charges must satisfy
\be \sum_{i=1}^{N_R} X_{\nu_{iR}} =3z,\hs \sum_{i=1}^{N_R} X^3_{\nu_{iR}} =3z^3.\label{anomaly}  \ee
Solving the equations in (\ref{anomaly}), we obtain the following results: there is no solution when $N_R=1$; there are only complex solutions when $N_R=2$, which are unacceptable; and there is an infinite number of real solutions when $N_R\geq 3$. Note that the $\nu_R$ results obtained here include those in the $B-L$ extension corresponding to the case of $z=-1$. Particularly considering the case of $N_R=3$ and requiring that at least two of three right-handed neutrinos be identical responsible for neutrino mass generation, we obtain two definite solutions. The first solution is $X_{\nu_{1,2,3R}}=z$, called conventional solution. The second solution is $X_{\nu_{1,2R}}=4z$ and $X_{\nu_{3R}}=-5z$, called alternative solution. Each of these solutions implies a realistic model, which will be investigated in the subsequent sections of the present work.

Last, but not least, for the remaining non-trivial anomalies, the $[SU(3)_C]^2U(1)_X$ anomaly is automatically canceled because the left and right chiral quarks have the same $X$ value, and the anomalies $[U(1)_Y]^2U(1)_X$ and $[U(1)_X]^2U(1)_Y$ also automatically vanish due to right-handed neutrinos having zero hypercharge. 

\section{\label{sec3}Conventional $U(1)_X$ model}

\subsection{Particle content}

In the conventional $U(1)_X$ model, three right-handed neutrinos are universal under $U(1)_X$. The fermion and scalar contents, as well as their quantum numbers under the gauge symmetry (\ref{gaugesymmetry}), are presented in Table \ref{tab1}, in which $z$ is arbitrarily nonzero. In the scalar sector, besides the SM Higgs doublet labeled as $\Ph$, a scalar singlet $\chi$ is necessarily presented to break $U(1)_X$, determining a residual symmetry $Z_2$, as well as generating appropriate right-handed neutrino masses through the coupling $\nu_R\nu_R\chi$. The scalar multiplets develop vacuum expectation values (VEVs),
 \be \langle \Ph\rangle=\begin{pmatrix}
0\\
\fr{v}{\sqrt2}\end{pmatrix},\hs \langle\chi\rangle=\fr{\La}{\sqrt2},\ee
satisfying $\La\gg v=246$ GeV for consistency with the SM. 
\begin{table}[h]
\bc
\begin{tabular}{l|cccccccc}
\hline\hline
Multiplets & $SU(3)_C$ & $SU(2)_L$ & $U(1)_Y$ & $U(1)_X$ & $Z_2$ \\ \hline 
$l_{aL}=(\nu_{aL},e_{aL})^T$ & $\bf 1$ & $\bf 2$ & $-\fr 1 2$ & $z$ & $+$ \\
$\nu_{aR}$ & $\bf 1$ & $\bf 1$ & $0$ & $z$ & $+$\\
$e_{aR}$ & $\bf 1$ & $\bf 1$ & $-1$ & $z$ & $+$\\
$q_{\al L}=(u_{\al L},d_{\al L})^T$ & $\bf 3$ & $\bf 2$ & $\fr 1 6$ & $-z$ & $+$\\
$u_{\al R}$ & $\bf 3$ & $\bf 1$ & $\fr 2 3$ & $-z$ & $+$\\
$d_{\al R}$ & $\bf 3$ & $\bf 1$ & $-\fr 1 3$ & $-z$ & $+$\\
$q_{3L}=(u_{3L},d_{3L})^T$ & $\bf 3$ & $\bf 2$ & $\fr 1 6$ & $z$ & $+$\\
$u_{3 R}$ & $\bf 3$ & $\bf 1$ & $\fr 2 3$ & $z$ & $+$\\
$d_{3 R}$ & $\bf 3$ & $\bf 1$ & $-\fr 1 3$ & $z$ & $+$\\
$\Ph=(\Ph_1^+,\Ph_2^0)^T$ & $\bf 1$ & $\bf 2$ & $\fr 1 2$ & $0$ & $+$\\
$\chi$ & $\bf 1$ & $\bf 1$ & $0$ & $-2z$ & $+$\\
$\xi$ & $\bf 1$ & $\bf 1$ & $0$ & $2z$ & $-$\\
$\eta$ & $\bf 1$ & $\bf 1$ & $0$ & $z$ & $-$\\
\hline\hline
\end{tabular}
\caption[]{\label{tab1}Matter content in the conventional $U(1)_X$ model.}
\ec
\end{table} 

\subsection{Matter parity and implication for single-component dark matter}
With the assumption $\La\gg v$, the gauge symmetry is broken as $SU(3)_C\otimes SU(2)_L\otimes U(1)_Y\otimes U(1)_X \stackrel{\La}\longrightarrow SU(3)_C\otimes SU(2)_L\otimes U(1)_Y\otimes R \stackrel{v}\longrightarrow SU(3)_C\otimes U(1)_Q\otimes R$,
where $Q=T_3+Y$ is as usual, while $R$ is a residual symmetry of $U(1)_X$ that conserves the $\chi$ vacuum. As being a $U(1)_X$ transformation, we have $R=e^{i\delta X}$, where $\delta$ is a transforming parameter. The vacuum conservation condition $R\langle \chi\rangle = \langle \chi\rangle $ implies $e^{i\delta (-2z)}=1$, or equivalently $\delta=k \pi/z$ for $k$ integer. Hence, the residual symmetry is 
\be R=e^{ik\pi X/z}=(-1)^{kX/z}.\ee 

It is clear that if $k = 0$, then $R = 1$ for all fields and every $z$, which is the identity transformation. Additionally, for $k\neq 0$, the relevant transformation $R=1$ is valid for all fields, given the minimal value of $|k|=2$. Hence, the residual symmetry $R$ is automorphic to a discrete group, such as
$\mathcal{Z}_2=\lbrace 1,g\rbrace$ with $g=(-1)^{X/z}$ and $g^2=1$. Because the spin parity, $p_s=(-1)^{2s}$, is always conserved by the Lorentz symmetry, we conveniently multiply the discrete group with the spin parity group $S=\lbrace1,p_s\rbrace$ with $p_s^2=1$, to perform a new group $\mathcal{Z}_2\otimes S$, which has an invariant discrete subgroup to be
\be Z_2=\lbrace 1,p\rbrace \ee
with $p=g\times p_s=(-1)^{X/z+2s}$ and $p^2=1$. Therefore, we decompose $\mathcal{Z}_2\otimes S\cong[(\mathcal{Z}_2\otimes S)/Z_2]\otimes Z_2$. Since $[(\mathcal{Z}_2\otimes S)/Z_2]= \lbrace\lbrace 1,p \rbrace,\lbrace g,p_s \rbrace\rbrace$ is conserved if $Z_2$ is conserved, we consider $Z_2$ to be a residual symmetry alternative to $\mathcal{Z}_2$. As usual, $Z_2$ has two one-dimensional irreducible representations, $\underline{1}$ according to $p = 1$ and $\underline{1}'$ according to $p = -1$. All the SM fields and new scalar singlet $\chi$ transform trivially under $Z_2$, indicated in the last column of Table \ref{tab1}.

That said, the conventional $U(1)_X$ model implies exactly a matter parity, $Z_2$. Because of the $Z_2$ conservation, the model can contain several scenarios for single-component dark matter, such as a single dark field to be either a dark (vectorlike) fermion, labeled $\xi$, or a dark scalar singlet, labeled $\eta$, which all transform nontrivially under $Z_2$, i.e., $p=-1$. In what follows, we extend the present model to include both $\xi$ and $\eta$ candidates. These fields and their simplest quantum numbers are collected to the last two rows of Table \ref{tab1}. Note that the candidate $\xi$ is vectorlike, and thus does not contribute to any gauge anomalies, while the candidate $\eta$ possesses vanishing VEV, i.e. $\langle\eta\rangle=0$, due to the conservation of $Z_2$.  

\subsection{Fermion mass and seesaw mechanism}

The spontaneous symmetry breaking will generate fermion masses through the Yukawa interactions, such as
\bea \mathcal{L} &\supset& h^e_{ab}\bar{l}_{aL} \Ph e_{bR}+h^\nu_{ab} \bar{l}_{aL} \tilde{\Ph} \nu_{bR}+\fr 1 2 f^\nu_{ab} \bar{\nu}^c_{aR} \nu_{bR}\chi\crn
&&+ h^d_{\al \beta} \bar{q}_{\al L} \Ph d_{\beta R}+ h^u_{\al \beta} \bar{q}_{\al L} \tilde{\Ph} u_{\beta R} +h^d_{33}\bar{q}_{3L} \Ph d_{3R}+ h^u_{33} \bar{q}_{3L} \tilde{\Ph} u_{3R} 
\crn
&&+ \fr{h^d_{\al 3}}{M} \bar{q}_{\al L} \Ph\chi d_{3 R}+ \fr{h^u_{3\al }}{M} \bar{q}_{3L} \tilde{\Ph}\chi^* u_{\al R} +\fr{h^d_{3\al }}{M}\bar{q}_{3L} \Ph\chi^* d_{\al R}+ \fr{h^u_{\al 3}}{M} \bar{q}_{\al L} \tilde{\Ph}\chi u_{3 R}\crn
&& + y_a\bar{\xi}_L\eta\nu_{aR}-m_\xi \bar{\xi}_L\xi_R +\mathrm{H.c.},\label{yu}\eea
where we have labeled $\tilde{\Ph}=i\sigma_2\Ph^*$ with $\sigma_2$ to be the second Pauli matrix, and $M$ is a new physics (or cutoff) scale that defines the effective interactions. Note that the couplings $h$, $f^\nu$, and $y$ are dimensionless, whereas $m_\xi$ has a mass dimension. 

From the above interactions, we obtain mass matrices for charged leptons, down-type quarks, and up-type quarks, which are given by
\bea [M_e]_{ab} &=& -h^e_{ab}\frac{v}{\sqrt2},\\
\left[M_d\right]_{\al\beta} &=& -h^d_{\al\beta}\frac{v}{\sqrt2}, \hs [M_d]_{33}=-h^d_{33}\frac{v}{\sqrt2}, \\
\left[M_{d}\right]_{\al 3} &=& -h^d_{\al 3}\fr{v\La}{2M}, \hs [M_d]_{3\beta}= -h^d_{3\beta}\frac{v\La}{2M},\\
\left[M_u\right]_{\al\beta} &=& -h^u_{\al\beta}\frac{v}{\sqrt2}, \hs [M_u]_{33}=-h^u_{33}\frac{v}{\sqrt2}, \\
\left[M_u\right]_{\al 3} &=& -h^u_{\al 3}\frac{v\La}{2M}, \hs [M_u]_{3\beta}= -h^u_{3\beta}\frac{v\La}{2M}.  \eea
Therefore, the small mixing between the third quark family and the first two quark families can be understood by either $h_{\al 3},h_{3\beta}<h_{\al\beta},h_{33}$ or $\La<M$. Diagonalizing these mass matrices, we get the masses of the relevant particles and the Cabibbo-Kobayashi-Maskawa matrix, as expected.

Concerning neutrinos, $\nu_{aL,R}$ achieve a Dirac mass via $h^\nu$ coupling, $[M_D]_{ab}=-h^\nu_{ab} \fr{v}{\sqrt{2}}$, while $\nu_{aR}$ obtains a Majorana mass via $f^\nu$ coupling, $[M_M]_{ab}=-f^\nu_{ab}\fr{\La}{\sqrt{2}}$. Thus, the total mass matrix of neutrinos takes the form, 
\be \mathcal{L}_{\mathrm{Yukawa}}\supset -\fr 1 2 (\bar{\nu}_L, \bar{\nu}^c_R)
\begin{pmatrix}
0 & M_D\\
M_D^T & M_M\end{pmatrix}
\begin{pmatrix}
\nu^c_L\\
\nu_R\end{pmatrix}+\mathrm{H.c.}\ee 
Because of $\La\gg v$, i.e. $M_M\gg M_D$, the active neutrinos $\sim\nu_L$ acquire a small mass via the canonical seesaw to be
\be [m_{\nu}]_{ab}=-[M_DM_M^{-1}M_D^T]_{ab}=(h^\nu)^2(f^\nu)^{-1}v^2/\sqrt2\La,\ee 
whereas the heavy neutrinos $\sim\nu_R$ obtain a large mass at the new physics scale $\La$. Additionally, if $h^\nu$ is very small, $h^\nu\sim 10^{-5}$, similar to electron Yukawa coupling, while  fixing $f^\nu\sim 0.6$, the observed value $m_\nu\sim 0.1$ eV requires $\La\sim 15$ TeV. Alternatively, given that $(h^\nu)^2/f^\nu\sim 1$, the model predicts $\La\sim 10^{14}$ TeV close to the grand unification scale.  

Lastly, since the VEV of dark scalar singlet vanishes due to $Z_2$ conservation, the dark fermion does not mix with right-handed neutrinos despite the term $y_a\bar{\xi}_L\eta\nu_{aR}$. Therefore, the dark fermion $\xi$ is a physical field by itself, with an arbitrary mass, $m_\xi$.

\subsection{Gauge sector}

When the gauge symmetry breaking takes place, the gauge bosons acquire masses via the kinetic terms of scalar fields, $\sum_S(D^\mu S)^\dagger (D_\mu S)$, where $S$ runs over scalar multiplets. The covariant derivative is defined as 
\be D_\mu = \pa_\mu + i g_s t_p G_{p\mu} + i g T_j A_{j\mu}+ i g_Y Y B_\mu + i g_X X C_\mu,\ee
in which $(g_s,g,g_Y, g_X)$, $(t_p, T_j, Y, X)$, and $(G_{p\mu}, A_{j\mu}, B_\mu, C_\mu)$ denote coupling constants, generators, and gauge bosons of $(SU(3)_C, SU(2)_L, U(1)_Y, U(1)_X)$ groups, respectively. 

The charged gauge bosons, $W^\pm$, take the form with corresponding mass,
\be W^\pm = \frac{1}{\sqrt2}(A_1\mp iA_2),\hs m_W = \fr{g^2v^2}{4}, \ee
which implies $v=246$ GeV. Because the SM scalar doublet $\Ph$ is not charged under $U(1)_X$, while the new scalar singlet $\chi$ is not charged under the SM gauge group, there is no mass mixing between the SM neutral gauge boson $Z$ and the new gauge boson $Z'$ coming from gauge symmetry breaking.\footnote{Kinetic mixing term between the two $U(1)$ gauge fields if imposed would cause a small effect, as suppressed.} It is straightforward to define the photon field $A$, the SM neutral gauge boson $Z$, and the new neutral gauge boson $Z'$, with their corresponding masses, as
\bea A &=& s_W A_3 + c_W B,\hs m_A=0,\\
Z &=& c_W A_3 - s_W B,\hs m^2_Z = \frac{g^2v^2}{4c^2_W},\\
Z' &=& C,\hs m^2_{Z'}=4g^2_Xz^2\La^2, \eea
where the Weinberg's angle is defined by $t_W \equiv \tan (\theta_W) = g_Y/g$, as usual.

\subsection{Scalar sector}

In presence of two scalar singlets $\chi$ and $\eta$ as well as the SM scalar doublet $\Ph$, the total scalar potential is given by 
\bea V &=& \mu^2_1 \Ph^\dagger \Ph + \mu^2_2\chi^*\chi + \mu^2_3\eta^*\eta + \la_1(\Ph^\dagger \Ph)^2+\la_2(\chi^*\chi)^2 +\la_3(\eta^*\eta)^2 \crn
&& + \la_4 (\Ph^\dagger \Ph)(\chi^*\chi) + \la_5 (\Ph^\dagger \Ph)(\eta^*\eta) + \la_6 (\chi^*\chi)(\eta^*\eta), \label{poten}\eea   
where $\la$'s is dimensionless, whereas $\mu$'s has a mass dimension. Necessary conditions for this potential to be bounded from below as well as to yield desirable vacuum structure are \be \la_{1,2,3} > 0,\hs \mu_{1,2}^2 < 0, \hs\mu_3^2 > 0, \hs |\mu_1|\ll |\mu_2|.\ee

To obtain the potential minimum and physical scalar
spectrum, we expand  
\bea \Ph &=& \begin{pmatrix}
\Ph^+_1\\
\frac{1}{\sqrt2}(v+S_1+iA_1)
\end{pmatrix} , \\
\chi &=& \frac{1}{\sqrt2}(\La +S_2+iA_2),\hs
\eta = \frac{1}{\sqrt2}(S_3+iA_3). \eea
Substituting them to the scalar potential (\ref{poten}), we get the potential minimum conditions,
\be \La^2=\frac{2(\la_4\mu_1^2-2\la_1\mu_2^2)}{4\la_1\la_2-\la_4^2},\hs v^2=\frac{2(\la_4\mu_2^2-2\la_2\mu_1^2)}{4\la_1\la_2-\la_4^2}. \ee
Further, we obtain physical scalar fields, such as
\be \Ph\simeq \begin{pmatrix}
G^+_W\\
\frac{1}{\sqrt2}(v+H+iG_{Z})
\end{pmatrix} , \hs \chi \simeq \frac{1}{\sqrt2}(\La + H'+iG_{Z'}), \ee 
where the mixing between two CP-even scalars $H=S_1$ and $H'=S_2$ is suppressed by $v/\La$, which has been neglected for simplicity. That said, $H$ is identical to the SM Higgs boson, while $H'$ is a new Higgs boson associated with $U(1)_X$ breaking. Their masses are 
\be m^2_H \simeq 2\la_1 v^2, \hs m^2_{H'} \simeq 2\la_2\La^2.\ee The CP-odd fields $G_W$, $G_{Z}$, and $G_{Z'}$ are massless Goldstone bosons, which are absorbed by $W$, $Z$, and $Z'$ gauge bosons, respectively.

The dark scalars $S_3$ and $A_3$ do not mix with the other scalars, because of the $Z_2$ conservation, and they are degenerate in mass. Therefore, they define a physical complex field $\eta$ with an arbitrary mass to be
\be m^2_\eta = \mu_3^2 + \fr 1 2 \la_5 v^2 + \fr 1 2 \la_6 \La^2. \ee

\section{\label{sec4}Alternative $U(1)_X$ model}
We would like to remind the reader that this model and the previous model share many common notations, which have similar properties and should be understood.

\subsection{Particle content}

In the alternative $U(1)_X$ model, only two right-handed neutrinos are universal under $U(1)_X$, which is different from the previous model. The particle content and their quantum numbers under the gauge symmetry in (\ref{gaugesymmetry}) are presented in Table \ref{tab2}. Concerning the scalar sector, in addition to the SM scalar doublet $\Ph$, the two scalar singlets $\chi_{1,2}$ are necessarily included to break $U(1)_X$, determining a residual symmetry $Z_2$ as well as generating appropriate Majorana masses for right-handed neutrinos through couplings $\nu_{\al R}\nu_{\al R}\chi_1$ and $\nu_{3R}\nu_{3R}\chi_2$. The scalar multiplets develop VEVs, such as
 \be \langle \Ph\rangle=\begin{pmatrix}
0\\
\fr{v}{\sqrt2}\end{pmatrix},\hs \langle\chi_1\rangle=\fr{\La_1}{\sqrt2}, \hs \langle\chi_2\rangle=\fr{\La_2}{\sqrt2},\ee
satisfying $\La_{1,2}\gg v=246$ GeV for consistency with the SM. 

It is stressed that since the charge assignment of $U(1)_X$, the neutrino mass is forbidden at tree level. We thus introduce two more scalars, namely, a doublet $\ph$ linking $l_{aL}$ to $\nu_{\al R}$ and a singlet $\eta$ linking $\ph$ to $\Ph\chi_1$ as well as to $\chi_2$. It is noted that both $\ph$ and $\eta$ are odd under a residual gauge symmetry $Z_2$ of $U(1)_X$ (see below) and cannot develop VEV due to the conservation of this $Z_2$. This implements a scotogenic mechanism for generating appropriate neutrino masses, as shown in Fig. \ref{fig1} \cite{Ma:2006km}.
\begin{table}[h]
\bc
\begin{tabular}{l|cccccccc}
\hline\hline
Multiplets & $SU(3)_C$ & $SU(2)_L$ & $U(1)_Y$ & $U(1)_X$ & $Z_2$ \\ \hline 
$l_{aL}=(\nu_{aL},e_{aL})^T$ & $\bf 1$ & $\bf 2$ & $-\fr 1 2$ & $z$ & $+$ \\
$\nu_{\al R}$ & $\bf 1$ & $\bf 1$ & $0$ & $4z$ & $-$\\
$\nu_{3R}$ & $\bf 1$ & $\bf 1$ & $0$ & $-5z$ & $+$\\
$e_{aR}$ & $\bf 1$ & $\bf 1$ & $-1$ & $z$ & $+$\\
$q_{\al L}=(u_{\al L},d_{\al L})^T$ & $\bf 3$ & $\bf 2$ & $\fr 1 6$ & $-z$ & $+$\\
$u_{\al R}$ & $\bf 3$ & $\bf 1$ & $\fr 2 3$ & $-z$ & $+$\\
$d_{\al R}$ & $\bf 3$ & $\bf 1$ & $-\fr 1 3$ & $-z$ & $+$\\
$q_{3L}=(u_{3L},d_{3L})^T$ & $\bf 3$ & $\bf 2$ & $\fr 1 6$ & $z$ & $+$\\
$u_{3 R}$ & $\bf 3$ & $\bf 1$ & $\fr 2 3$ & $z$ & $+$\\
$d_{3 R}$ & $\bf 3$ & $\bf 1$ & $-\fr 1 3$ & $z$ & $+$\\
$\Ph=(\Ph_1^+,\Ph_2^0)^T$ & $\bf 1$ & $\bf 2$ & $\fr 1 2$ & $0$ & $+$\\
$\chi_1$ & $\bf 1$ & $\bf 1$ & $0$ & $-8z$ & $+$\\
$\chi_2$ & $\bf 1$ & $\bf 1$ & $0$ & $10z$ & $+$\\
$\ph=(\ph_1^0,\ph_2^-)^T$ & $\bf 1$ & $\bf 2$ & $-\fr 1 2$ & $-3z$ & $-$\\
$\eta$ & $\bf 1$ & $\bf 1$ & $0$ & $-5z$ & $-$\\
\hline\hline
\end{tabular}
\caption[]{\label{tab2}Matter content in the alternative $U(1)_X$ model.}
\ec
\end{table} 

\subsection{\label{TCDM}Matter parity and implication for two-component dark matter}

The spontaneous symmetry breaking of the gauge symmetry (\ref{gaugesymmetry}) down to $SU(3)_C\otimes SU(2)_L\otimes U(1)_Y\otimes R$ is implemented by the new scalar singlets, $\chi_{1,2}$. Then, the SM gauge group is spontaneously broken to the low energy theory by the SM Higgs doublet, $\Ph$, as usual. Because $R$ is a residual symmetry of $U(1)_X$ that conserves both the vacua of $\chi_1$ and $\chi_2$, the transformation $R=e^{i\delta X}$ satisfies simultaneously $R\langle \chi_{1,2}\rangle = \langle \chi_{1,2}\rangle $, where $\delta$ is a transforming parameter. This leads to $e^{i\delta (-8z)}=1$ and $e^{i\delta (10z)}=1$, implying $\delta=k \pi/z$ for $k$ integer, thus $R=e^{ik\pi X/z}=(-1)^{kX/z}$. Similar to the previous model, the residual symmetry $R$ is automorphic to the discrete group $\mathcal{Z}_2=\lbrace 1,g\rbrace$ with $g=(-1)^{X/z}$ and $g^2=1$. Also, since the spin parity $p_s=(-1)^{2s}$ is always conserved, we multiply $p_s$ with $g$ to form $p=g\times p_s=(-1)^{X/z+2s}$, which performs a discrete group $Z_2=\lbrace 1,p\rbrace$ with $p^2=1$ to be a residual symmetry instead of $\mathcal{Z}_2$, for convenience. Under $Z_2$, all the SM fields, $\nu_{3R}$, and $\chi_{1,2}$ are even, whereas $\nu_{\al R}$, $\ph$, and $\eta$ are all odd, as presented in the last column of Table \ref{tab2}.

The model under consideration is of particular interest, with a novel implication for two-component dark matter. Indeed, because of the conservation of $Z_2$, the lightest field of odd fields $\nu_{\al R}, \ph, \eta$ is stabilized, impossibly decayed to normal fields, providing a dark matter candidate. In addition, the third right-handed neutrino $\nu_{3R}$ with charge $X=-5z$ do not couple to any SM particles due to the gauge invariance and hence it reveals accidentally an alternative candidate for dark matter. In other words, a promising scenario for two-component dark matter, which consists of the lightest odd-field and $\nu_{3R}$, is hinted.

\subsection{Gauge sector}

In the current model, since the SM scalar doublet $\Ph$ does not transform under $U(1)_X$ and that the new scalar singlets $\chi_{1,2}$ do not transform under $SU(2)_L\otimes U(1)_Y$, there is no tree-level mixing between the gauge bosons $Z$ and $Z'$, similar to the previous model. The SM gauge bosons $W^\pm, A, Z$ and new gauge boson $Z'$ with their masses are given by
\bea W^\pm &=& \frac{1}{\sqrt2}(A_1\mp iA_2),\hs m^2_W = \fr{g^2v^2}{4} ,\\
 A &=& s_W A_3 + c_W B,\hs m_A=0,\\
Z &=& c_W A_3 - s_W B,\hs m^2_Z = \frac{g^2v^2}{4c^2_W},\\
Z' &=& C,\hs m^2_{Z'}= g^2_Xz^2(64\La^2_1+100\La^2_2), \eea
where the Weinberg's angle is defined by $t_W = g_Y/g$, as usual.

\subsection{Scalar sector relevant for symmetry breaking}
As shown above, two new scalar singlets $\chi_{1,2}$ break $U(1)_X$ to $Z_2$, and then the SM scalar doublet $\Ph$ breaks the $SU(2)_L\otimes U(1)_Y$ electroweak symmetry down to $U(1)_Q$, as usual. The scalar potential consisting of $\Ph, \chi_{1,2}$ is given by
\bea  V &\supset& \mu_0^2\Ph^\dag\Ph + \mu^2_1\chi_1^*\chi_1 + \mu^2_2\chi_2^*\chi_2 + \la_0(\Ph^\dag\Ph)^2 + \la_1(\chi_1^*\chi_1)^2 + \la_2(\chi_2^*\chi_2)^2 \crn 
&&+ (\Ph^\dag\Ph)[\la_3(\chi_1^*\chi_1) + \la_4(\chi_2^*\chi_2)] + \la_5(\chi_1^*\chi_1)(\chi_2^*\chi_2).\label{poten2} \eea
The necessary conditions for this potential to be bounded from below as well as yielding a desirable vacuum structure are
\be \la_{0,1,2}>0,\hs \mu^2_{0,1,2}<0,\hs |\mu_0|\ll|\mu_{1,2}|.\ee 
Note that the scalar potential in Eq. (\ref{poten2}) actually has a global $U(1)$ symmetry in addition to the gauge symmetry. Consequently, there is a physical Goldstone boson in the particle spectrum after the spontaneous symmetry breaking, which was discussed in \cite{Rothstein:1992rh}.

Expanding the scalar fields around their VEVs as
\bea \Ph &=& \begin{pmatrix}
\Ph^+_1\\
\frac{1}{\sqrt2}(v+S_1+iA_1)
\end{pmatrix} , \\
\chi_1 &=& \frac{1}{\sqrt2}(\La_1 +S_2+iA_2),\hs
\chi_2 = \frac{1}{\sqrt2}(\La_2+S_3+iA_3), \eea
and substituting these expressions into the scalar potential (\ref{poten2}), we obtain the following potential minimum conditions:
\bea 2\mu_0^2+2\la_0v^2+\la_3\La_1^2+\la_4\La_2^2 &=&0,\\
2\mu_1^2+\la_3v^2+2\la_1\La_1^2+\la_5\La_2^2 &=&0,\\
2\mu_2^2+\la_4v^2+\la_5\La_1^2+2\la_2\La_2^2 &=&0. \eea
Further, we obtain physical scalar fields, such as
\bea \Ph &\simeq& \begin{pmatrix}
G^+_W\\
\frac{1}{\sqrt2}(v+H+iG_Z)
\end{pmatrix} , \\
\chi_1 &\simeq& \frac{1}{\sqrt2}(\La_1 + H_1+i\mathcal{A}),\hs
\chi_2 \simeq \frac{1}{\sqrt2}(\La_2 +H_2+iG_{Z'}), \eea
where we have assumed $\La_2\gg \La_1\gg v$ and neglected the mixing among three CP-even scalars for simplicity. That said, the assumption $\La_2\gg \La_1$ would lead to an attractive scenario of two-component fermion dark matter (cf. Subsection \ref{TCDMP} below). Above, $H$ is identical to the SM Higgs boson, whereas $H_{1,2}$ are the new Higgs bosons. Their masses are approximately given by 
\be  m^2_H\simeq 2\la_0 v^2, \hs m^2_{H_1}\simeq 2\la_1 \La_1^2, \hs m^2_{H_2}\simeq 2\la_2 \La_2^2.\ee
Additionally, $G_W$, $G_Z$, and $G_{Z'}$ are the massless Goldstone bosons absorbed by $W$, $Z$, and $Z'$ gauge bosons, respectively, whereas $\mathcal{A}$ is a physical Goldstone boson as predicted.

The presence of a physical Goldstone boson, in principle, poses some issues for the model. However, the physical Goldstone boson $\mathcal{A}$ here is probably safe because it does not directly couple to the SM particles except for the Higgs boson whose couplings are well controlled by the parameters in the potential, and it decouples from the thermal bath in the early universe. Indeed, the $\la_3(\Ph^\dag\Ph)(\chi_1^*\chi_1)$ term allows the SM Higgs to decay into a pair of Goldstone bosons, inducing a tree-level contribution to partial width as $\Gamma_{H\to\mathcal{A}\mathcal{A}}\sim \la_3^2v^2/32\pi m_H$. Taking $m_H\simeq 125.25$ GeV, the full width of the Higgs boson to be $3.2$ MeV, and the branching ratio for invisible decay models must be less than $13\%$ \cite{ParticleDataGroup:2022pth}, we obtain a constraint on the coupling to be $|\la_3|\lesssim 0.009$. On the other hand, following the discussion in Ref. \cite{Weinberg:2013kea}, since the $H_{1,2}$ and $Z'$ bosons that couple to $\mathcal{A}$ are heavy, their interactions with $\mathcal{A}$ decouple at sufficiently high temperature in the early universe. Concerning the interaction between $\mathcal{A}$ and the SM fermions, the ratio $R$ between the rate of collisions of $\mathcal{A}$ with fermion $f$ and the expansion rate of the universe, at a temperature $T$, is roughly given by $R\sim \la_3^2 m^2_f (kT)^5m_{\mathrm{P}}/(m_{H_1}m_H)^4$ \cite{Weinberg:2013kea}, where $m_f$ is the mass of the fermion $f$, $m_{\mathrm{P}}$ is the Planck mass, and $k$ is the Boltzmann constant. The decoupling of $\mathcal{A}$ from thermal equilibrium occurs when the ratio is equal to unity, i.e. $R\sim 1$. Taking $m_{\mathrm{P}}\simeq 1.22 \times 10^{19}$ GeV \cite{ParticleDataGroup:2022pth}, and assuming $m_f=m_\tau=1.77686$ GeV (the tauon mass), we obtain a decoupling temperature as $T_\mathrm{d}\sim 9.54$ GeV, which is consistent with the condition $T>m_\tau$, given that $k\sim 1$, $|\la_3|\sim 0.009$, and $m_{H_1}=1$ TeV. In addition, it is clear that this decoupling temperature is far above the neutrino decoupling temperature approximately a few MeV, as well as the QCD phase transition temperature about 200 MeV. Therefore, the contribution of $\mathcal{A}$ to the density of radiation in the Universe that is usually parametrized by the effective neutrino number $N_{\mathrm{eff}}$ is given by \be \Delta N_{\mathrm{eff}} = \frac{4}{7}\left(\frac{h_\gamma^{\mathrm{BBN}}}{h_\gamma^{\mathrm{dec}}}\right)^{4/3} \lesssim \frac{4}{7}\left(\frac{10.75}{60}\right)^{4/3}\simeq 0.06,\ee where $h_\gamma^{\mathrm{BBN}}$ and $h_\gamma^{\mathrm{dec}}$ are the entropy degrees of freedom of the plasma at the time of big bang nucleosynthesis and of the $\mathcal{A}$ decoupling, respectively. This result is in agreement with the current bound on $N_{\mathrm{eff}}$, i.e. $N_{\mathrm{eff}}=2.99\pm 0.17$~\cite{Planck:2018vyg}. All imply that if $|\la_3|\lesssim 0.009$, $\mathcal{A}$ decouples from the thermal bath in the early universe around $\mathcal{O}(1)$ GeV temperature, and this does not cause serious problems in particle physics or cosmology. 

\subsection{Fermion mass and scotogenic mechanism}

When the scalar multiplets develop VEVs, the fermions acquire masses through Yukawa interactions, such as
\bea \mathcal{L} &\supset& h^e_{ab}\bar{l}_{aL} \Ph e_{bR}+ h^d_{\al \beta} \bar{q}_{\al L} \Ph d_{\beta R}+ h^u_{\al \beta} \bar{q}_{\al L} \tilde{\Ph} u_{\beta R} +h^d_{33}\bar{q}_{3L} \Ph d_{3R}+ h^u_{33} \bar{q}_{3L} \tilde{\Ph} u_{3R} 
\crn
&&+ \fr{h^d_{\al 3}}{M^2} \bar{q}_{\al L} \Ph\chi_1^*\chi_2^* d_{3 R}+ \fr{h^u_{3\beta }}{M^2} \bar{q}_{3L} \tilde{\Ph}\chi_1\chi_2 u_{\beta R} +\fr{h^d_{3\beta }}{M^2}\bar{q}_{3L} \Ph\chi_1\chi_2 d_{\beta R}+ \fr{h^u_{\al 3}}{M^2} \bar{q}_{\al L} \tilde{\Ph}\chi_1^*\chi_2^* u_{3 R}\crn
&&+h^\nu_{a\bet} \bar{l}_{aL} \ph \nu_{\bet R}+\fr 1 2 f^\nu_{\al\bet} \bar{\nu}^c_{\al R} \nu_{\bet R}\chi_1 +\fr 1 2 f^\nu_{33} \bar{\nu}^c_{3R} \nu_{3R}\chi_2 + \mathrm{H.c.}\label{yu2}\eea
The right-handed neutrinos capture large Majorana masses at $\La_{1,2}$ scales, 
\be m_{\nu_{1R}}=-f^\nu_{11}\frac{\La_1}{\sqrt2}, \hs m_{\nu_{2R}}=-f^\nu_{22}\frac{\La_1}{\sqrt2},\hs m_{\nu_{3R}}=-f^\nu_{33}\frac{\La_2}{\sqrt2}, \ee
assuming that $f^\nu_{\al\bet}$ is flavor diagonal, without loss of generality, i.e. $\nu_{1,2,3R}$ are physical fields by themselves. Whereas, the charged leptons, down-type quarks, and up-type quarks obtain mass matrices at the weak scale, namely
\bea [M_e]_{ab} &=& -h^e_{ab}\frac{v}{\sqrt2}, \\{}[M_d]_{\al\beta} &=& -h^d_{\al\beta}\frac{v}{\sqrt2}, \hs [M_d]_{33}=-h^d_{33}\frac{v}{\sqrt2}, \\
\left[M_{d}\right]_{\al 3} &=& -h^d_{\al 3}\fr{v\La_1\La_2}{2\sqrt2 M^2}, \hs [M_d]_{3\beta}= -h^d_{3\beta}\fr{v\La_1\La_2}{2\sqrt2 M^2},  \\{}[M_u]_{\al\beta} &=& -h^u_{\al\beta}\frac{v}{\sqrt2}, \hs [M_u]_{33}=-h^u_{33}\frac{v}{\sqrt2}, \\
\left[M_u\right]_{\al 3} &=& -h^u_{\al 3}\fr{v\La_1\La_2}{2\sqrt2 M^2}, \hs [M_u]_{3\beta}= -h^u_{3\beta}\fr{v\La_1\La_2}{2\sqrt2 M^2}.  \eea
These mass matrices provide appropriate masses for the corresponding particles after diagonalization, given that $h_{\al 3},h_{3\beta}<h_{\al\beta},h_{33}$ and/or $\La_{1,2}< M$. 

\begin{figure}[h]
\includegraphics[scale=1]{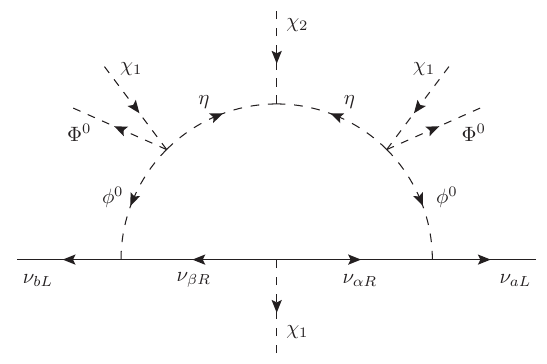}
 \caption[]{\label{fig1}Radiative generation of neutrino mass through dark matter.}
\end{figure}

Differing from the previous model, the active neutrinos in the current model obtain radiative Majorana masses through the scotogenic mechanism with the $Z_2$-odd fields $\nu_{\al R}, \ph$, and $\eta$, as described in Fig. \ref{fig1}. 
To get the relevant physical scalars, we consider the scalar potential related to $\ph$ and $\eta$, which is given by
\bea V &\supset& \mu^2_3\ph^\dag\ph + \mu^2_4\eta^*\eta + \la_6(\ph^\dag\ph)^2 + \la_7(\eta^*\eta)^2 + (\Ph^\dag\Ph)[\la_8(\ph^\dag\ph) + \la_9(\eta^*\eta)] \crn 
&&+ (\chi_1^*\chi_1)[\la_{10}(\ph^\dag\ph) + \la_{11}(\eta^*\eta)]+ (\chi_2^*\chi_2)[\la_{12}(\ph^\dag\ph) + \la_{13}(\eta^*\eta)]\crn
&&+ \la_{14}(\ph^\dag\ph)(\eta^*\eta) + \la_{15}(\Ph^\dag\ph)(\ph^\dag\Ph)+\mu(\chi_2\eta\eta+\mathrm{H.c.})+\la(\Ph\ph\eta\chi^*_1+\mathrm{H.c.}). \eea
The necessary conditions for this potential to be bounded from below as well as yielding the trivial vacua for both $\ph$ and $\eta$ ensuring the $Z_2$ symmetry are $\mu^2_{3,4}>0$ and $\la_{6,7}>0$. Expanding $\ph=((S_4+iA_4)/\sqrt2,\ph_2^-)^T$ and $\eta=(S_5+iA_5)/\sqrt2$, then substituting them into the scalar potential, we obtain
\bea V &\supset& \fr 1 2 \begin{pmatrix} S_4 & S_5 \end{pmatrix} \begin{pmatrix}  M^2_1 & -\fr 1 2 \la v \La_1\\
-\fr 1 2 \la v \La_1 & M^2_2+\sqrt{2}\mu \La_2 \end{pmatrix} \begin{pmatrix} S_4\\ S_5\end{pmatrix}\crn
&&+\fr 1 2 \begin{pmatrix} A_4 & A_5 \end{pmatrix} \begin{pmatrix} M^2_1 & \fr 1 2 \la v \La_1\\
\fr 1 2 \la v \La_1 & M^2_2-\sqrt{2}\mu \La_2 \end{pmatrix} \begin{pmatrix} A_4\\ A_5\end{pmatrix}, \eea 
where $M^2_1=(2\mu^2_3+\la_{10}\La^2_1+\la_{12}\La^2_2+\la_8v^2)/2$ and $M^2_2=(2\mu^2_4+\la_{11}\La^2_1+\la_{13}\La^2_2+\la_9v^2)/2$. Defining two mixing angles $\theta_{R,I}$ via the tan functions as 
\be t_{2\theta_R}=\fr{\la v \La_1}{M^2_1-M^2_2-\sqrt{2}\mu \La_2},\hs t_{2\theta_I}=\fr{ \la v \La_1}{M^2_2-M^2_1-\sqrt{2}\mu \La_2},\ee
we get the physical fields
\bea \mathcal{S}_1 &=& c_{\theta_R} S_4 - s_{\theta_R} S_5,\hs \mathcal{S}_2=s_{\theta_R} S_4 + c_{\theta_R} S_5, \\ 
\mathcal{A}_1 &=& c_{\theta_I} A_4 - s_{\theta_I} A_5,\hs \mathcal{A}_2 =s_{\theta_I} A_4 + c_{\theta_I} A_5, \eea 
with respective masses,
\bea m^2_{\mathcal{S}_1} &\simeq& M^2_1-\fr{\fr 1 4 \la^2 v^2 \La^2_1}{M^2_2-M^2_1+\sqrt{2}\mu \La_2},\hs m^2_{\mathcal{S}_2}\simeq M^2_2+\sqrt{2}\mu \La_2+\fr{\fr 1 4 \la^2 v^2 \La^2_1}{M^2_2-M^2_1+\sqrt{2}\mu \La_2},\\
m^2_{\mathcal{A}_1} &\simeq& M^2_1-\fr{\fr 1 4 \la^2 v^2 \La^2_1}{M^2_2-M^2_1-\sqrt{2}\mu \La_2},\hs m^2_{\mathcal{A}_2}\simeq M^2_2-\sqrt{2}\mu \La_2+\fr{\fr 1 4 \la^2 v^2 \La^2_1}{M^2_2-M^2_1-\sqrt{2}\mu \La_2}, \eea
in which the approximations come from the conditions $\La_{1,2}\gg v$. Lastly, the radiative neutrino mass matrix is given by \cite{VanDong:2023thb} 
\bea (m_\nu)_{ab} &=& \fr{h^\nu_{a\al}h^\nu_{b\al}m_{\nu_{\al R}}}{32\pi^2}\left(\fr{c^2_{\theta_R} m^2_{\mathcal{S}_1}}{m^2_{\mathcal{S}_1}-m^2_{\nu_{\al R}}}\ln \fr{m^2_{\mathcal{S}_1}}{m^2_{\nu_{\al R}}}-\fr{c^2_{\theta_I}m^2_{\mathcal{A}_1}}{m^2_{\mathcal{A}_1}-m^2_{\nu_{\al R}}}\ln \fr{m^2_{\mathcal{A}_1}}{m^2_{\nu_{\al R}}}\right.\crn
&&\left.+\fr{s^2_{\theta_R}m^2_{\mathcal{S}_2}}{m^2_{\mathcal{S}_2}-m^2_{\nu_{\al R}}}\ln \fr{m^2_{\mathcal{S}_2}}{m^2_{\nu_{\al R}}}-\fr{s^2_{\theta_I}m^2_{\mathcal{A}_2}}{m^2_{\mathcal{A}_2}-m^2_{\nu_{\al R}}}\ln \fr{m^2_{\mathcal{A}_2}}{m^2_{\nu_{\al R}}}\right).\eea
With the assumption $\La_2\gg\La_1\gg v$, both the $\mathcal{S}_1, \mathcal{A}_1$ mass splitting $(m^2_{\mathcal{S}_1}-m^2_{\mathcal{A}_1})/m^2_{\mathcal{S}_1,\mathcal{A}_1}$ and the mixing angles $\theta^2_{R,I}$ are proportional to $\la^2v^2\La^2_1/\La^4_2$. Hence, the observed neutrinos obtain small Majorana masses, $m_\nu\sim (h^\nu)^2\la^2v^2\La^3_1/32\pi^2\La_2^4$, as expected. For example, if $h^\nu\sim 0.01$, $\la\sim 0.1$, the observed value $m_\nu\sim 0.1$ eV requires $\La_1\sim 1$ TeV and $\La_2\sim 6.5$ TeV.

In the subsequent sections, we will give a combined analysis of phenomenology of both the models in the question.  

\section{\label{sec5}FCNC and collider bounds}

\subsection{FCNCs}
 
In both models under consideration, the quark families transform differently under $U(1)_X$. This causes the FCNCs at the tree level, which are mediated only by the new neutral gauge boson $Z'$. Indeed, from the fermion kinetic terms, $\sum_F \bar{F}i\gamma^\mu D_\mu F$, where $F$ runs over fermion multiplets, we extract the couplings of $Z'$ to quarks induced by $X$-charge,
\be \mathcal{L}\supset -g_XX_{q_a}\bar{q}_a\gamma^\mu q_a Z'_\mu = zg_X(\bar{q}_1\gamma^\mu q_1+\bar{q}_2\gamma^\mu q_2-\bar{q}_3\gamma^\mu q_3)Z'_\mu, \label{fcnc}\ee
where $q_a$ labels quarks of either up-types ($u_a$) or down-types ($d_a$), and note that all the quarks are vector-like under $X$. Let us change to the mass basis by the transformation, $q_{aL,R}=[V_{q {L,R}}]_{ai}q_{iL,R}$, where $q_i$ denotes mass eigenstates of either up-types, $u_i=(u,c,t)$, or down-types, $d_i=(d,s,b)$, and $V_{q {L,R}}$ are the unitary matrices satisfying
$V_{u L}^\dag M_u V_{u R}=\mathrm{diag}(m_u,m_c,m_t)$ and $V_{d L}^\dag M_d V_{d R}=\mathrm{diag}(m_d,m_s,m_b)$. We rewrite (\ref{fcnc}) as
\be \mathcal{L}\supset zg_X\bar{q}_i\gamma^\mu q_iZ'_\mu-2zg_X[V_{q L}^*]_{3i}[V_{q L}]_{3j}\bar{q}_{iL}\gamma^\mu q_{jL}Z'_\mu+ (L\to R),\ee
where $i,j$ are summed. It is easy to see that the last two terms give rise to FCNCs for $i\neq j$. Integrating $Z'$ out, we get related effective interactions,
\be \mathcal{H}^{\mathrm{eff}}_{\mathrm{FCNC}}= \fr{L^2_{ij}}{m^2_{Z'}} (\bar{q}_{iL}\ga^\mu q_{jL})^2+ 2(LR)+(RR),\label{effinter}\ee 
where $L_{ij}=-2zg_X[V^*_{qL}]_{3i}[V_{qL}]_{3j}$, and the last two terms differ from the first one only in chiral structures. Further, the above effective couplings are approximated as 
\be \fr{L^2_{ij}}{m^2_{Z'}}\simeq \fr{1}{\bar{\La}^2}([V^*_{qL}]_{3i}[V_{qL}]_{3j})^2,\ee 
where we have defined $\bar{\La}^2\equiv\La^2$ for the conventional $U(1)_X$ model, while $\bar{\La}^2\equiv 16\La_1^2+25\La_2^2$ for the alternative $U(1)_X$ model. 

The effective interactions in (\ref{effinter}) contribute to the relevant neutral-meson mixing amplitudes. Assuming the new physics effects dominantly comes from the first term in (\ref{effinter}), the existing data imply corresponding bounds, 
\be \fr{L^2_{12}}{m^2_{Z'}}<\left(\fr{1}{10^4\ \mathrm{TeV}}\right)^2,\hs \fr{L^2_{13}}{m^2_{Z'}}<\left(\fr{1}{500\ \mathrm{TeV}}\right)^2,\hs \fr{L^2_{23}}{m^2_{Z'}}<\left(\fr{1}{100\ \mathrm{TeV}}\right)^2,\ee according to $K^0$-$\bar{K}^0$ mixing, $B^0_d$-$\bar{B}^0_d$ mixing, and $B^0_s$-$\bar{B}^0_s$ mixing, respectively \cite{UTfit:2007eik,Isidori:2010kg}. Additionally, aligning quark mixing to the down sector, i.e. $V_{uL}=1$, the CKM matrix is given by $V_{\mathrm{CKM}}=V_{uL}^\dag V_{dL}=V_{dL}$, thus $[V_{dL}]_{31}=0.00857$, $[V_{dL}]_{32}=0.04110$, and $[V_{dL}]_{33}=0.999118$~\cite{ParticleDataGroup:2022pth}, which lead to the constraints, 
\be \bar{\La}\gtrsim 3.52\ \mathrm{TeV},\hs \bar{\La}\gtrsim 4.28\ \mathrm{TeV},\hs \bar{\La}\gtrsim 4.11\ \mathrm{TeV},\label{fcncbound1}\ee corresponding to the mentioned meson mixing systems.

Generically, the last two terms of (\ref{effinter}) also contribute to the neutral-meson mixing amplitudes through switching on the right-handed quark mixing matrix, $V_{qR}$. Since $V_{qR}$ is left arbitrarily as in the standard model, we assume $V_{qR}\simeq V_{qL}$ for simplicity and also the reason that the current gauge symmetry that contains $B,L$ may obey a left-right symmetry at high energy. The contribution arising from all effective interactions in (\ref{effinter}) to the mass splitting in $K^0$-$\bar{K}^0$ mixing system, where $(q_i,q_j)=(d,s)$, is given by  
\bea \Delta m_{K}|_{\mathrm{NP}} &=& 2\mathrm{Re}\langle K^0|\mathcal{H}^{\mathrm{eff}}_{\mathrm{FCNC}}|\bar{K}^0\rangle\crn
 &=& 2 \mathrm{Re}\langle K^0|\fr{L^2_{12}}{m^2_{Z'}}(\bar{d}_L \ga^\mu s_L)^2+2\fr{L_{12}R_{12}}{m^2_{Z'}}(\bar{d}_L \ga^\mu s_L)(\bar{d}_R \ga_\mu s_R) +\fr{R^2_{12}}{m^2_{Z'}}(\bar{d}_R \ga^\mu s_R)^2|\bar{K}^0\rangle.\eea
The related hadronic matrix elements are  
\bea && \langle K^0|(\bar{d}_L \ga^\mu s_L)^2|\bar{K}^0\rangle= \langle K^0|(\bar{d}_R \ga^\mu s_R)^2|\bar{K}^0\rangle=\fr 1 3 m_K f^2_K,\\ 
&& \langle K^0|(\bar{d}_L \ga^\mu s_L)(\bar{d}_R \ga_\mu s_R)|\bar{K}^0\rangle = -\fr 1 2 \left[\fr 1 2 +\fr 1 3 \left(\fr{m_K}{m_d+m_s}\right)^2\right] m_K f^2_K,\eea 
which have been determined in the vacuum insertion approximation using PCAC \cite{Gabbiani:1996hi}, in agreement with \cite{Langacker:2000ju}. Hence,
\be \Delta m_{K}|_{\mathrm{NP}} \simeq \fr{2m_K f^2_K}{3m^2_{Z'}} \mathrm{Re}\left\{L^2_{12}-\left[\fr 3 2+ \left(\fr{m_K}{m_d+m_s}\right)^2\right]L_{12}R_{12}+R^2_{12}\right\}.\label{NPK}\ee  
Similarly for $B^0_{d,s}$-$\bar{B}^0_{d,s}$ mixings with $(q_i,q_j)=(d,b)$ and $(s,b)$, respectively, we have 
\bea \Delta m_{B_d}|_{\mathrm{NP}} &\simeq& \fr{2m_{B_d} f^2_{B_d}}{3m^2_{Z'}} \mathrm{Re}\left\{ L^2_{13}-\left[\fr 3 2+ \left(\fr{m_{B_d}}{m_d+m_b}\right)^2\right]L_{13}R_{13}+R^2_{13}\right\},\label{NPBd}\\ 
\Delta m_{B_s}|_{\mathrm{NP}} &\simeq& \fr{2m_{B_s} f^2_{B_s}}{3m^2_{Z'}} \mathrm{Re}\left\{ L^2_{23}-\left[\fr 3 2+ \left(\fr{m_{B_s}}{m_s+m_b}\right)^2\right]L_{23}R_{23}+R^2_{23}\right\}.\label{NPBs}\eea 

Note that the theoretical predictions of meson mass differences can be decomposed as
\be \Delta m_{K,B_d,B_s} = \Delta m_{K,B_d,B_s}|_{\mathrm{SM}} + \Delta m_{K,B_d,B_s}|_{\mathrm{NP}}, \ee
where the first part/term comes from the SM contribution \cite{Buras:2016dxz,Lenz:2019lvd},
\bea \Delta m_{K}|_{\mathrm{SM}} &=& 3.074\times 10^{-12} \text{ MeV}, \\
\Delta m_{B_d}|_{\mathrm{SM}} &=& (3.574\pm 0.191)\times 10^{-10} \text{ MeV}, \\
\Delta m_{B_s}|_{\mathrm{SM}} &=& (1.2355\pm 0.0566)\times 10^{-8} \text{ MeV}, \eea
while the second part/term is due to the new physics contribution as derived in (\ref{NPK}), (\ref{NPBd}), and (\ref{NPBs}). These theoretical predictions will be compared with the experimental values as summarized in \cite{ParticleDataGroup:2022pth}, namely
\bea \Delta m_{K}|_{\mathrm{exp}} &=& (3.484\pm 0.006)\times 10^{-12} \text{ MeV}, \\
\Delta m_{B_d}|_{\mathrm{exp}} &=& (3.334\pm 0.013)\times 10^{-10} \text{ MeV}, \\
\Delta m_{B_s}|_{\mathrm{exp}} &=& (1.1693\pm 0.0004)\times 10^{-8} \text{ MeV}. \eea
However, due to the presence of long-distance effects in $\Delta m_K$, the uncertainties in this system are very large. We therefore require the theory to produce the data for kaon mass difference within $\pm 30\%$, i.e., $0.7<\Delta m_K|_{\mathrm{SM}}/\Delta m_K|_{\mathrm{exp}}<1.3$, or equivalently \be -0.3<\frac{\Delta m_K|_{\mathrm{NP}}}{\Delta m_{K}|_{\mathrm{exp}}}<0.3, \ee 
in agreement with \cite{Buras:2015kwd}. For the $B_{d,s}$-meson mass differences, the SM predictions are more accurate than that of kaon, so we can calculate the $2\sigma$ ranges by combining quadrature of the relative errors in the SM predictions and measurements as
\be 0.83<\frac{\Delta m_{B_d}|_{\mathrm{exp}}}{\Delta m_{B_d}|_{\mathrm{SM}}}<1.04,\hs 0.86<\frac{\Delta m_{B_s}|_{\mathrm{exp}}}{\Delta m_{B_s}|_{\mathrm{SM}}}<1.03, \ee
and then we obtain the following constraints,
\be -0.17<\frac{\Delta m_{B_d}|_{\mathrm{NP}}}{\Delta m_{B_d}|_{\mathrm{SM}}}<0.04,\hs -0.14<\frac{\Delta m_{B_s}|_{\mathrm{NP}}}{\Delta m_{B_s}|_{\mathrm{SM}}}<0.03. \ee

By using the following input parameters \cite{ParticleDataGroup:2022pth,FlavourLatticeAveragingGroupFLAG:2021npn,Bazavov:2017lyh}, in units of MeV,
\bea m_d &=& 4.67,\hs   m_s = 93.4,\hs  m_b = 4180,\\
m_K &=& 497.611,\hs   m_{B_d} = 5279.66,\hs  m_{B_s} = 5366.92,\\
f_K &=& 155.7,\hs  f_{B_d} = 190.5,\hs  f_{B_s} = 230.7, \eea
we obtain the constraints
\be \bar{\La}\gtrsim 4.91\ \mathrm{TeV},\hs \bar{\La}\gtrsim 12.97\ \mathrm{TeV},\hs \bar{\La}\gtrsim 14.14\ \mathrm{TeV},\label{fcncbound2}\ee
corresponding to the $K^0$-$\bar{K}^0$, $B^0_d$-$\bar{B}^0_d$, and $B^0_s$-$\bar{B}^0_s$ mixing systems. These lower bounds are generally more stringent and are quite larger than the those determined in (\ref{fcncbound1}).\footnote{At low energy, $Z'$ may also contribute to atomic parity violation but would be safely suppressed by the current FCNC bound (see, \cite{Dong:2006cn}, for a discussion).}

Alternatively, one could align the quark mixing to the up quark sector by assuming $V_{dL}=1$ and then $V_{\text{CKM}}=V_{uL}^\dag V_{dL}=V_{uL}^\dag$. Hence, the current $D^0$-$\bar{D}^0$ mixing data implies 
\be \fr{[V_{\text{CKM}}]_{13}[V^*_{\text{CKM}}]_{23}}{\bar{\La}} < \fr{1}{10^3\ \mathrm{TeV}},\ee
given that the first term in Eq. (\ref{effinter}) dominantly contributes to the relative meson mixing amplitude \cite{UTfit:2007eik,Isidori:2010kg}. Taking $[V_{\text{CKM}}]_{13}=0.00369$ and $[V^*_{\text{CKM}}]_{23}=0.04182$ \cite{ParticleDataGroup:2022pth}, we obtain a constraint $\bar{\La}\gtrsim 0.15$ TeV, which is much weaker than those obtained in Eq. (\ref{fcncbound1}). Further, the mass splitting in this system has the form similar to Eq. (\ref{NPK}), i.e.
\be\Delta m_{D}|_{\mathrm{NP}} \simeq \fr{2m_D f^2_D}{3m^2_{Z'}} \mathrm{Re}\left\{ L^2_{12}-\left[\fr 3 2+ \left(\fr{m_D}{m_u+m_c}\right)^2\right]L_{12}R_{12}+R^2_{12}\right\},\ee
if all the terms in Eq. (\ref{effinter}) for contribution \cite{Golowich:2007ka}. For this scenario, using the input values given in MeV as \cite{ParticleDataGroup:2022pth,Bazavov:2017lyh},
\bea \Delta m_D|_{\mathrm{exp}} &=& (6.56237\pm 0.76353)\times 10^{-12},\\
m_D &=& 1864.84,\hs f_D = 211.6,\hs m_u = 2.16,\hs m_c = 1270,\eea
and requiring $-0.3<\Delta m_D|_{\mathrm{NP}}/\Delta m_D|_{\mathrm{exp}}<0.3$ as done in the $K^0$-$\bar{K}^0$ mixing system, we get a corresponding constraint as $\bar{\La}\gtrsim 1.04$ TeV, which is also much weaker than those obtained in Eq. (\ref{fcncbound2}).

\subsection{LEP-II}

The LEP experiments have made measurements in electron-positron collisions with collision center-of-mass energy ranging from the $Z$ pole (LEP-I) up to 209 GeV (LEP-II), providing stringent constraints on $Z'$ boson \cite{Carena:2004xs,ALEPH:2013dgf}. One of the processes searched at the LEP experiments is the four-fermion contact interaction mediated by heavy $Z'$ boson, which can be parameterized by the following effective Lagrangian \cite{Eichten:1983hw},
\bea
\mathcal{L}_{\text{eff}}&=&\frac{1}{1+\delta_{ef}}\frac{1}{m^2_{Z'}}\sum_{i,j=L,R}C^{Z'}_{e_i}C^{Z'}_{f_j}\bar{e}_i\gamma_\mu e_i \bar{f}_j\gamma^\mu f_j,
\eea
where $\delta_{ef}=1(0)$ for $f=e$ ($f\neq e$). From the relevant data of the LEP-II experiment \cite{ALEPH:2013dgf}, we impose a constraint on this contact interaction as
\be
\frac{2\sqrt{2\pi}m_{Z'}}{\sqrt{(C_{e_L}^{Z'})^2+(C_{e_R}^{Z'})^2}}\gtrsim 24.6\ \ \text{TeV}.
\ee
Note that the electron is vector-like under $U(1)_X$, i.e. $C^{Z'}_{e_L}=C^{Z'}_{e_R}=C^{Z'}_e=-zg_X$, for both models under consideration. Therefore, we get a lower bound of the new physics scale, $\bar{\La}\gtrsim 3.47$ TeV, which is independent of $g_X$ and $z$ and is quite smaller than the bounds obtained from FCNCs, shown in Eq. (\ref{fcncbound2}).

\subsection{LHC}
At the LHC experiment, the new gauge boson $Z'$ can be resonantly produced via the fusion processes $\bar{q}q\to Z'$ in which $q$ indicates the SM quarks, and this boson subsequently decays into the SM fermions as well as the exotic particles if kinetically allowed. Furthermore, the most significant decay channel of $Z'$ is given by $Z'\to l^+l^-$ with $l=e,\mu,\tau$ because of well-understood backgrounds \cite{ATLAS:2019erb} and also that it signifies a $Z'$ having both couplings to lepton and quark like ours. Using the narrow width approximation, the cross section for the
relevant process takes the form,
\be
\sigma(pp\rightarrow Z'\rightarrow l^+l^-) \simeq \sigma(pp\rightarrow Z')\text{Br}(Z'\rightarrow l^+l^-).\label{sigppll}\ee
The first factor is the production cross section of the $Z'$ boson, which can be estimated as
\be\sigma(pp\rightarrow Z')=\sum_{q}L_{q\bar{q}}(m^2_{Z'})\hat{\sigma}(\bar{q}q\rightarrow Z'),\ee
where $L_{q\bar{q}}$ is the parton luminosity,
\be
L_{q\bar{q}}(m^2_{Z'})=\int^1_{\fr{m^2_{Z'}}{s}}\frac{dx}{xs}\left[f_q(x,m^2_{Z'})f_{\bar{q}}\left(\frac{m^2_{Z'}}{xs},m^2_{Z'}\right)+f_q\left(\frac{m^2_{Z'}}{xs},m^2_{Z'}\right)f_{\bar{q}}(x,m^2_{Z'})\right],
\ee
with $\sqrt{s}$ to be the collider center-of-mass energy and $f_{q(\bar{q})}(x,m^2_{Z'})$ to be the parton distribution function of the quark $q$ (antiquark $\bar{q}$) evaluated at the scale $m^2_{Z'}$, while the partonic cross section is approximated as $\hat{\sigma}(\bar{q}q\rightarrow Z')\simeq \frac{\pi}{3}(C^{Z'}_q)^2$. The last factor in Eq. (\ref{sigppll}) is the branching ratio of $Z'$ decaying into the lepton pair $l^+l^-$,
\be\text{Br}(Z'\rightarrow l^+l^-)= \frac{\Gamma(Z'\rightarrow l^+l^-)}{\Gamma_{Z'}}, \ee 
where the partial decay width of $Z'$ boson is approximately given by
\be \Gamma(Z'\rightarrow l^+l^-)\simeq \frac{m_{Z'}}{12\pi}(C_l^{Z'})^2,\ee
while the total $Z'$ decay width is estimated as \bea
\Gamma_{Z'}&\simeq&\frac{m_{Z'}}{12\pi}\sum_f N_C(f)(C_f^{Z'})^2+\frac{m_{Z'}}{8\pi}(C^{Z'}_{\nu_L})^2\crn
&&+\frac{m_{Z'}}{24\pi}(C^{Z'}_{\nu_R})^2\sum^3_{i=1}\left(1-\frac{4m^2_{\nu_{iR}}}{m^2_{Z'}}\right)^{3/2}\Theta\left(\frac{m_{Z'}}{2}-m_{\nu_{iR}}\right)\crn
&&+\frac{m_{Z'}}{12\pi}(C_\xi^{Z'})^2\left(1-\frac{4m^2_\xi}{m^2_{Z'}}\right)^{3/2}\Theta\left(\frac{m_{Z'}}{2}-m_\xi\right)\crn
&&+\frac{m_{Z'}}{48\pi}(zg_X)^2\left(1-\frac{4m^2_{\eta}}{m^2_{Z'}}\right)^{3/2}\Theta\left(\frac{m_{Z'}}{2}-m_{\eta}\right),
\eea
for the conventional $U(1)_X$ model, assuming $m_{Z'}<m_{H'}$, and 
\bea
\Gamma_{Z'}&\simeq&\frac{m_{Z'}}{12\pi}\sum_f N_C(f)(C_f^{Z'})^2+\frac{m_{Z'}}{8\pi}(C^{Z'}_{\nu_L})^2+\frac{m_{Z'}}{12\pi}(C^{Z'}_{\nu_{\al R}})^2+\frac{m_{Z'}}{48\pi}(-8zg_X)^2\crn
&&+\frac{m_{Z'}}{24\pi}(C^{Z'}_{\nu_{3R}})^2\left(1-\frac{4m^2_{\nu_{3R}}}{m^2_{Z'}}\right)^{3/2}\Theta\left(\frac{m_{Z'}}{2}-m_{\nu_{3R}}\right),
\eea
for the alternative $U(1)_X$ model, assuming that $Z'$ is lighter than the new Higgs bosons $H_2,\mathcal{S}_{1,2}$, and $\mathcal{A}_{1,2}$. Above, $f$ denotes the SM charged fermions, $N_f$ is the color number of the fermion $f$, $\Theta$ is the step function, and the related couplings are given by
\be C^{Z'}_{e,\mu,\tau} = -C^{Z'}_{u,c,d,s}=C^{Z'}_{t,b}=C^{Z'}_{\nu_L} =  C^{Z'}_{\nu_R} =\fr 1 2 C^{Z'}_\xi = \fr 1 4 C^{Z'}_{\nu_{\al R}}= -\fr 1 5 C^{Z'}_{\nu_{3R}} = -zg_X.\label{couplings}\ee 

\begin{figure}[h]
\includegraphics[scale=0.42]{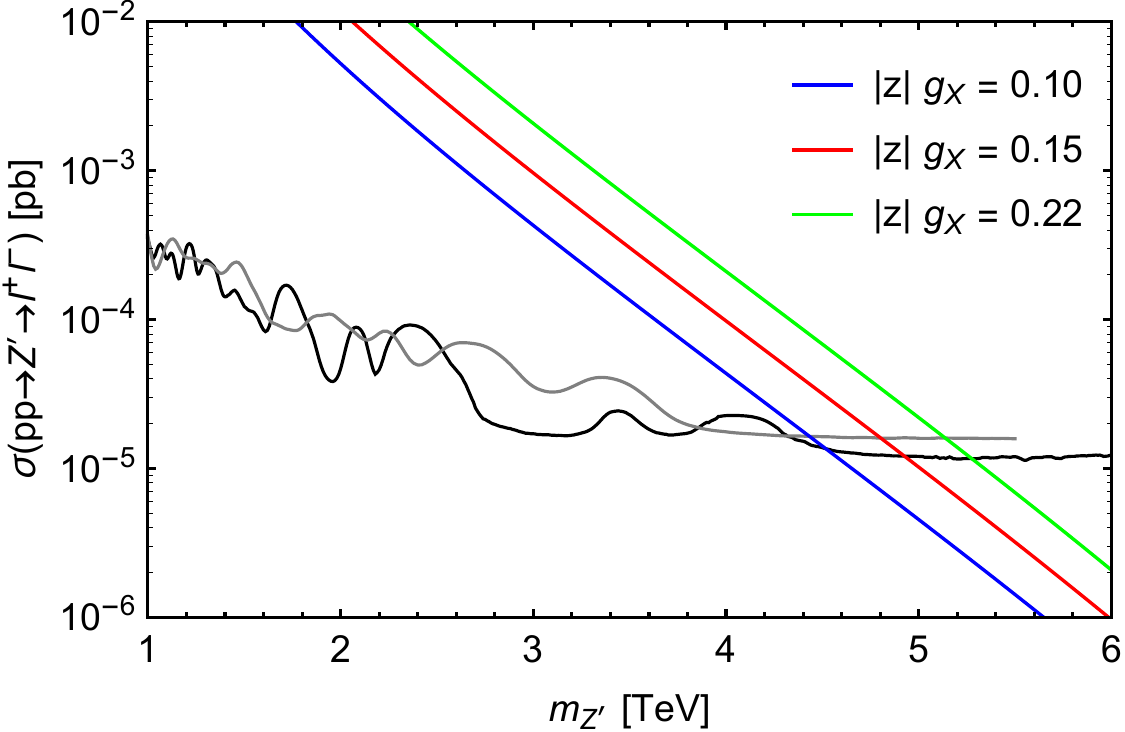}
\includegraphics[scale=0.42]{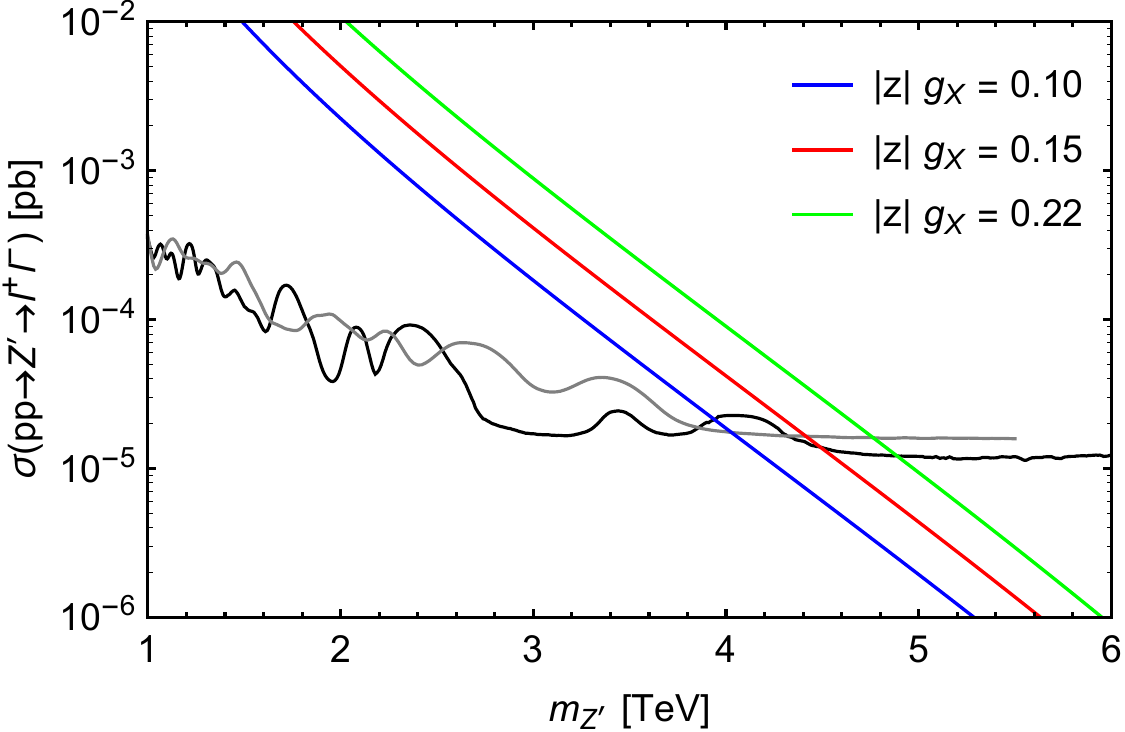}
 \caption[]{\label{fig2}Dilepton production cross-section as a function of the $Z'$ boson mass for various values of the $|z|g_X$ product. The left (right) panel corresponds to the conventional (alternative) $U(1)_X$ model. The black (gray) curve shows the upper bound on the cross-section obtained by the ATLAS 2019 results for $\Gamma/m=3\%$  \cite{ATLAS:2019erb} (the CMS 2019 results \cite{CMS:2021ctt} for $\Gamma/m=0.6\%$).}
\end{figure}

Employing the MSTW2008 parton distribution functions \cite{Martin:2009iq} and setting $\sqrt{s}=13$ TeV, we plot the cross-section for the relevant process as a function of the $Z'$ boson mass, for various values of the product $|z|g_X$, which are shown in Fig. \ref{fig2}. The left panel corresponds to the convention $U(1)_X$ model, assuming $m_{\nu_{iR}}=2m_\xi/3=4m_\eta/3=m_{Z'}/3$, while the right panel corresponds to the alternative $U(1)_X$ model, given that $m_{\nu_{3R}}= m_{Z'}/2$. In addition, in each of these panels, we include the upper limits on the cross section of this process observed by ATLAS \cite{ATLAS:2019erb} and CMS \cite{CMS:2021ctt} experiments. In the left (right) panel, the lower bounds on the $Z'$ boson mass are $4.5, 4.9$, and $5.3$ ($4.0, 4.5$, and $4.9$) TeV according to $|z|g_X = 0.1, 0.15$, and $0.22$.

\begin{figure}[h]
\includegraphics[scale=0.42]{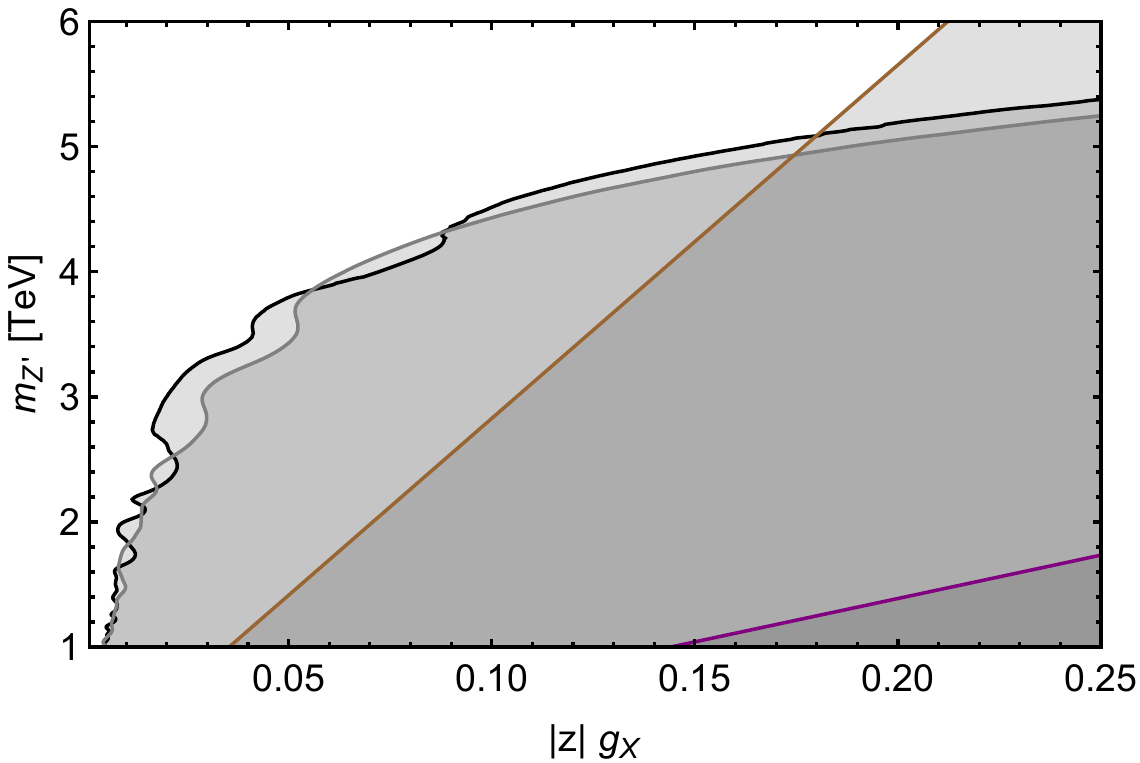}
\includegraphics[scale=0.42]{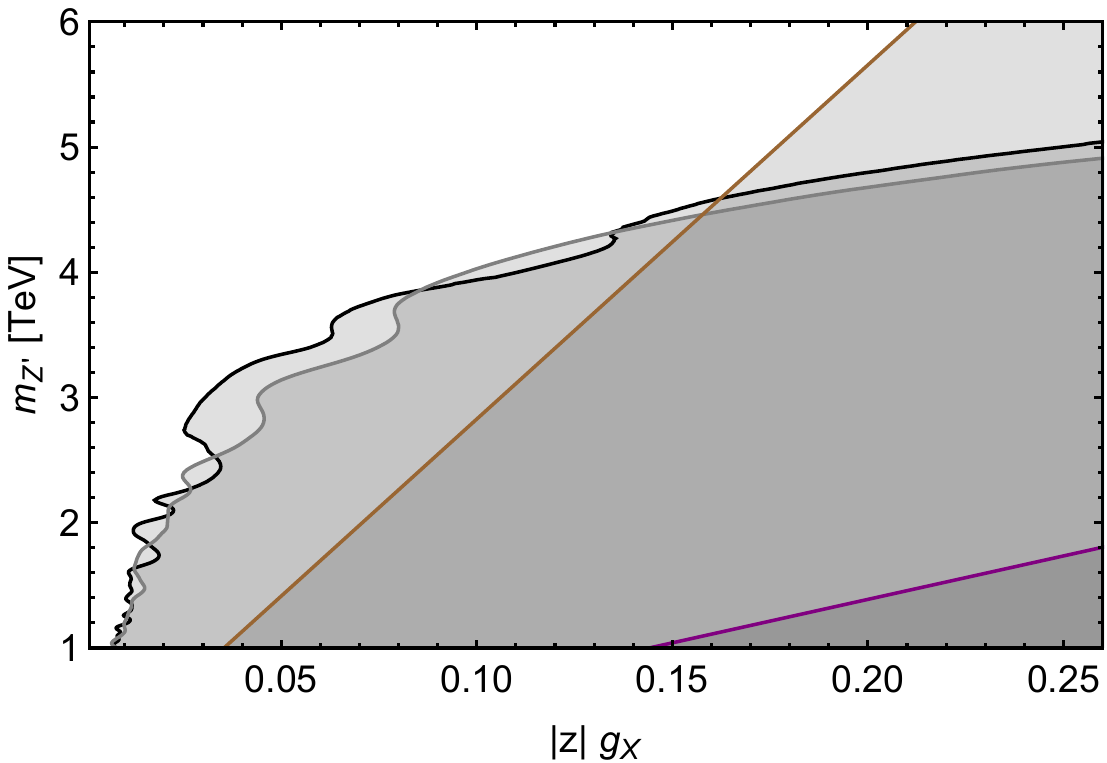}
 \caption[]{\label{fig3}The left (right) panel corresponds to the conventional (alternative) $U(1)_X$ model. For each panel, the black (gray), brown, and purple lines denote the lower bounds on the $Z'$ boson mass obtained from the $Z'$ boson search by the ATLAS (CMS) collaboration, the FCNCs, and the LEP-II dilepton signal constraint, respectively. The parameter space according to shaded regions is excluded.}
\end{figure}

In Fig. \ref{fig3}, the lower bounds of $Z'$ boson mass that obtain from the ATLAS (CMS) are described by the black (gray) curve. The left (right) panel corresponds to the convention (alternative) $U(1)_X$ model. For comparison, in each of these panels, we also add the lower bounds on the $Z'$ boson mass which come from the FCNCs (brown line) and the LEP-II (purple line). The available regions for the $Z'$ boson mass lie above these four lines. It is easy to see that the constraint from the ATLAS and CMS (FCNCs) is the strongest if the coupling strength $|z|g_X$ is smaller (larger) than about $0.18$ for the convention $U(1)_X$ model and $0.16$ for the alternative $U(1)_X$ model.

\section{\label{sec6}Dark matter phenomenology}

\subsection{Single-component dark matter in the conventional $U(1)_X$ model}

As shown, the dark matter candidate in the model, i.e., the lightest field of $\xi$ and $\eta$, can possess an arbitrary mass, which is stabilized due to the conservation of the $Z_2$ residual symmetry, $p=(-1)^{X/z+2s}$. Additionally, the observed light neutrino masses require the new physics scale $\La$ either in the TeV region or close to the grand unification scale, depending on magnitude of the Yukawa couplings, $h^\nu$ and $f^\nu$. Hence, the model predicts two different scenarios of dark matter production. \ben \item If the new physics scale is at the TeV region, the freeze-out mechanism works and defines not only the dark matter relic density through its annihilation into the SM particles as well as others when kinetically allowed, but also the dark matter nature to be a weakly interacting massive particle (WIMP). \item If the new physics scale is of order the grand unification or around, the dark matter may be asymmetrically produced from the $CP$-violation decay of the right-handed neutrinos, via complex couplings $y_a\bar{\xi}_L\eta \nu_{aR}$, similar to the leptogenesis mechanism for lepton asymmetry generation. \een However, only the first scenario is considered in this work due to the potential discovery of new physics at the LHC.

\subsubsection{Dark matter as a fermion $\xi$}

Assume that the fermion $\xi$ is lighter than the scalar $\eta$. Then $\xi$ is stabilized, responsible for dark matter. Because $\xi$ is charged under $U(1)_X$, its pair annihilation into the SM particles and others (if kinetically allowed) in the early universe proceeds dominantly through $s$-channel exchange diagrams mediated by the new gauge boson $Z'$. Hence, the dark matter annihilation cross section times relative velocity is estimated as
\bea \langle\sigma v_{\text{rel}}\rangle_{\xi\xi^c\to\text{all}} &\simeq &  \frac{(C^{Z'}_\xi)^2m^2_\xi}{\pi(4 m_\xi^2-m^2_{Z'})^2}\left[\sum_f N_C(f) (C^{Z'}_f)^2+\frac{3(C^{Z'}_{\nu_L})^2}{2}\right.\crn
&&\left.+\fr{(C^{Z'}_{\nu_R})^2}{2}\sum^3_{i=1}\left(1-\frac{m_{\nu_{iR}}^2}{4m^2_\xi}\right)\left(1-\frac{m_{\nu_{iR}}^2}{m^2_\xi}\right)^{1/2}\Theta(m_\xi-m_{\nu_{iR}})\right],\eea
where $f$ denotes the SM charged fermions, $N_f$ is the color number of the fermion $f$, $\Theta$ is the step function, and the related couplings are shown in Eq. (\ref{couplings}). Hence, the relic abundance of fermion dark matter is given by
\be \Om_\xi h^2\simeq \fr{0.1 \text{ pb}}{\langle\sigma v_{\text{rel}}\rangle_{\xi\xi^c\to\text{all}}}.\ee 

\begin{figure}[h!]
\includegraphics[scale=0.42]{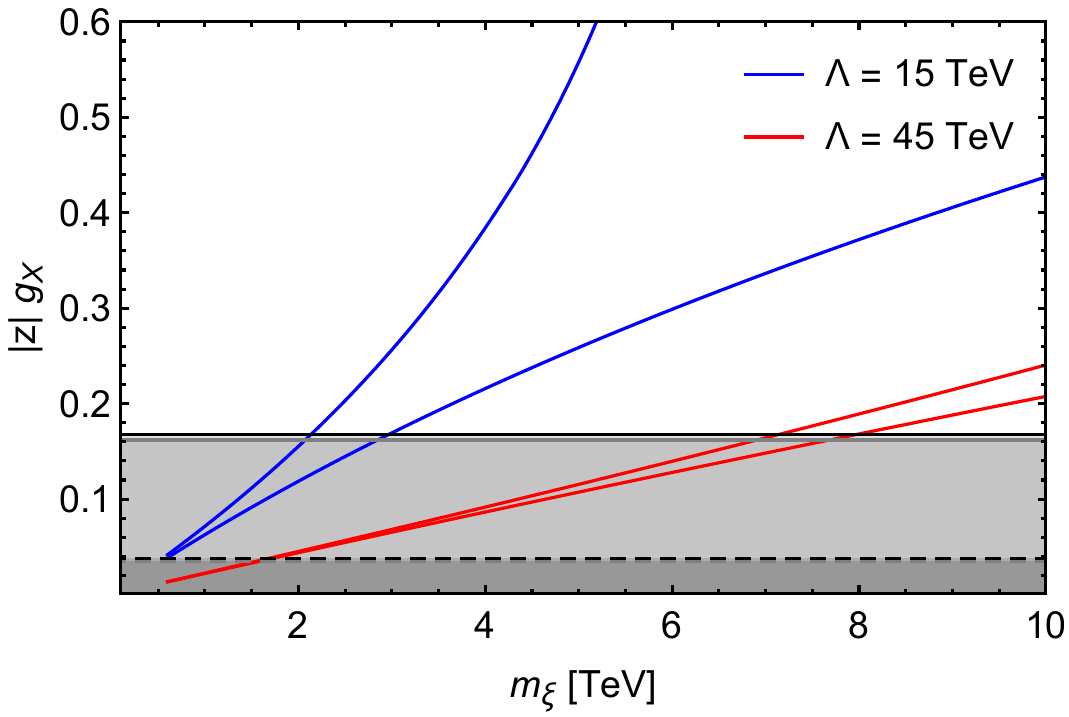}
\includegraphics[scale=0.42]{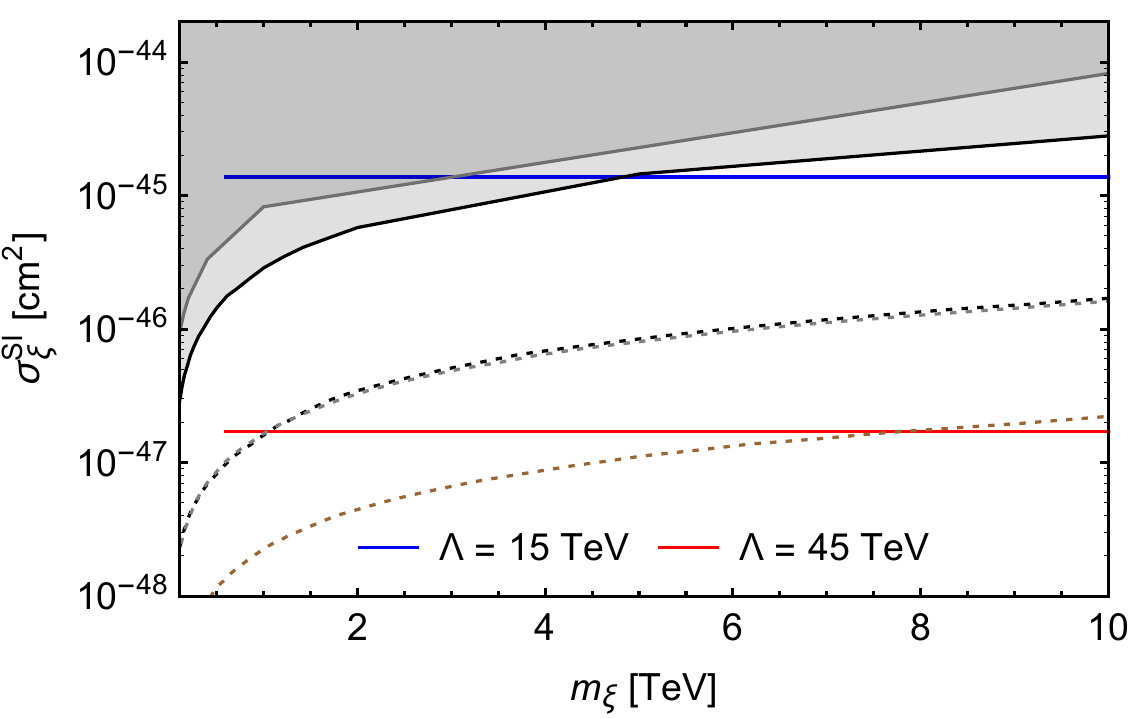}
 \caption[]{\label{fig4}In the left panel, we make the blue (red) contours of the correct dark matter relic density at $\La=15$ (45) TeV. In the right panel, we make the blue (red) lines of the SI scattering cross section of the dark fermion on a nucleon at $\Lambda=15~(45)$ TeV. The shaded regions are excluded by the current experiments. The dotted curves are the projected bounds.}
\end{figure}

Let $\La=15\ (45)$ TeV satisfy the limits from the FCNCs and the LEP-II (a future projected bound), and take $m_{\nu_{iR}}=m_{Z'}/3$. According to $\La=15 ~(45)$ TeV, we make the blue (red) contours of the correct dark matter relic density, i.e. $\Om_\xi h^2\simeq 0.12$ \cite{Planck:2018vyg}, as shown in the left panel in Fig. \ref{fig4}. Added to these density contours are the black (dashed black) and gray (dashed gray) lines corresponding to $|z|g_X\simeq 0.168~(0.038)$ and $|z|g_X\simeq 0.162~(0.035)$ which are extracted from the ATLAS and CMS limits at $\La=15 ~(45)$ TeV, respectively, where notice that the shaded regions are excluded. From here, we obtain the lower bound for fermion dark matter mass to be $m_\xi\gtrsim 2.1 ~(1.7)$ TeV for $\La=15 ~(45)$ TeV.

Besides the correct dark matter relic density, the fermion candidate is constrained by the direct dark matter detection experiments which measure the scattering cross section of the dark matter on nucleons in target nucleus. At the microscopic level, the scattering of the dark matter on nucleons can be described by effective interactions between the dark matter and the SM quarks. In the model under consideration, such interactions are dominantly contributed by $t$-channel diagrams by the new gauge boson $Z'$ exchange, given by
\be\mathcal{L}^{\text{eff}}_{\xi-\text{quark}}=\frac{1}{m^2_{Z'}}C^{Z'}_\xi C^{Z'}_q(\bar{\xi}\gamma^\mu \xi)(\bar{q}\gamma_\mu q), \ee
where $q=u,d$. Note that both $\xi$ and the SM quarks are vector-like under $U(1)_X$. Hence, we obtain the spin-independent (SI) scattering cross section of $\xi$ on a nucleon, labeled as $\mathcal{N}$ with corresponding mass $m_{\mathcal{N}}$, such as \cite{Belanger:2008sj}
\be \sigma^{\text{SI}}_\xi=\frac{9\mu^2_{\xi\mathcal{N}}}{4\pi}\frac{1}{m^4_{Z'}}(C^{Z'}_\xi)^2(C^{Z'}_u)^2\simeq 1.377\times 10^{-45} \left(\frac{m_\mathcal{N}}{1\text{ GeV}}\right)^2\left(\frac{15\text{ TeV}}{\La}\right)^4 \text{ cm}^2,\ee
which is independent of the product $|z|g_X$. Above, $\mu_{\xi\mathcal{N}}=m_\xi m_\mathcal{N}/(m_\xi+m_\mathcal{N})\simeq m_\mathcal{N}$ is the reduced mass of the dark matter-nucleon system.

Taking $m_{\mathcal{N}}\simeq 1$ GeV and $\Lambda=15$ TeV as before, the model predicts $\sigma^{\text{SI}}_\xi\simeq 1.377\times 10^{-45} \text{ cm}^2$ as described by the blue line in the right panel in Fig. \ref{fig4}, which agrees with the current bound from XENON1T \cite{XENON:2018voc} (LZ \cite{LZ:2022lsv}) that is described by the gray (black) curve, if dark fermion mass satisfies $m_\xi\gtrsim 3.0~(4.9)$ TeV. These lower bounds for the fermion dark matter mass are generally stricter than that obtained from the correct density constraint. Further, we include the projected bounds from upcoming direct detection experiments such as XENONnT \cite{XENON:2020kmp}, LZ \cite{LZ:2018qzl}, and DARWIN \cite{DARWIN:2016hyl}, which are respectively described by the gray, black, and brown dotted curves. It is clear that these bounds constrain the new physics scale and/or the dark fermion mass to higher values. Take an example, the model predicts $\sigma^{\text{SI}}_\xi\simeq 1.7\times 10^{-47} \text{ cm}^2$ for $\Lambda=45$ TeV as descried by the red line, which agrees with the projected bounds if dark fermion mass satisfies $m_\xi\gtrsim 7.9$ TeV.

\subsubsection{Dark matter as a scalar $\eta$}

Alternatively, we consider a possibility that the dark scalar singlet $\eta$ is lighter than the dark fermion $\xi$. So, $\eta$ is stabilized, responsible for dark matter. For simplicity, we assume that $\eta$ is lighter than $H'$, $Z'$, and $\nu_{iR}$, and thus it annihilates only to SM particles in the early universe. Such annihilation may proceed through a contact interaction with the SM Higgs fields as well as $H$, $H'$, and $Z'$ portals. However, the contribution of $Z'$ portal to dark matter observables is quite similar to the case of the fermion dark matter above, whereas the $H'$ portal does not contribute to direct dark matter detection. Hence, we do not consider these two portals here, given that the contributions of these two portals are negligible, by imposing appropriate parameters. With the contact interaction and $H$ portal, the dark matter pair annihilation into the SM particles is given by the following dominant channels:
\be \eta^*\eta\to HH, W^+W^-, ZZ, \ee
 yielding the dark matter annihilation cross section times relative velocity to be
\be \langle\sigma v_{\text{rel}}\rangle_{\eta^*\eta\to\text{SM SM}}\simeq \frac{\la_5^2}{16\pi m^2_\eta}. \ee
Assuming that the scalar dark matter $\eta$ provides a correct relic density, i.e. $\Om_\eta h^2\simeq 0.1\text{ pb}/\langle\sigma v_{\text{rel}}\rangle_{\eta^*\eta\to\text{SM SM}}\simeq 0.12$, we derive the dark matter mass at the TeV regime, i.e.
\be m_\eta\simeq |\la_5|\times 3.05 \ \text{TeV}. \ee

Concerning the direct dark matter detection, the effective Lagrangian that describes the interaction between $\eta$ and the SM quarks induced by $t$-channel exchange diagrams of the new Higgs boson $H$ is given by 
\be \mathcal{L}^{\text{eff}}_{\eta-\text{quark}}= \frac{\la_5 m_q}{m^2_H}\eta^*\eta \bar{q}q\ee
with $q$ denotes ordinary quarks. From this Lagrangian, we acquire the SI scattering cross section of $\eta$ on a nucleon, such as \cite{Barger:2008qd}
\be\sigma^{\text{SI}}_\eta =\left(\fr{\la_5}{2\sqrt{\pi}}\frac{\mu_{\eta \mathcal{N}}}{m^2_H}\frac{m_\mathcal{N}}{m_\eta}C_\mathcal{N}\right)^2,  \ee
where $\mu_{\eta \mathcal{N}}=m_\eta m_\mathcal{N}/(m_\eta+ m_\mathcal{N})\simeq m_\mathcal{N}$, and
\be C_\mathcal{N}=\frac{2}{9}+\frac{1}{A}\sum_{q=u,d,s}\left[\left(Z-\frac{2}{9}A\right) f^p_{q}+(A-Z)f^n_{q}\right], \ee
in which $Z$ is the nucleus charge, $A$ is the total number of nucleons in the nucleus, and $f^{p(n)}_q$ are the scalar couplings of the nucleon to $q=u,d,s$. The values of $f^{p(n)}_q$ are given by \cite{Junnarkar:2013ac,Hoferichter:2015dsa}
\be f^{p(n)}_u \simeq 0.0208(0.0189), \hs f^{p(n)}_d \simeq 0.0411(0.0451), \hs f^{p(n)}_s \simeq 0.043(0.043). \ee
Taking $A=131$, $Z=54$, $m_{\mathcal{N}}\simeq 1$ GeV, and $m_H\simeq 125.25$ GeV \cite{ParticleDataGroup:2022pth}, we estimate
\be\sigma^{\text{SI}}_\eta\simeq 1.259\times 10^{-45} \left(\frac{|\la_5|\times 3.05\text{ TeV}}{m_\eta}\right)^{2} \text{ cm}^2.\ee
This result implies that the scalar dark matter $\eta$ with a mass at TeV regime, $m_\eta\simeq |\la_5|\times 3.05$ TeV, yields both the correct relic density and the current exclusion limit in direct dark matter detection experiments \cite{XENON:2018voc,PandaX-II:2020oim,LUX:2016ggv}.

\subsection{\label{TCDMP}Two-component dark matter in the alternative $U(1)_X$ model}

As shown in the subsection \ref{TCDM}, the alternative $U(1)_X$ model provides two candidates for dark matter simultaneously, the one being $\nu_{3R}$ accidentally stabilized and the remainder to be the lightest of $Z_2$-odd fields ($\nu_{\al R}$, $\ph$, $\eta$) stabilized by $p=(-1)^{X/z+2s}$. With the assumption $\La_2\gg\La_1$ as discussed in the previous subsections, the $Z_2$-odd physical scalars are generally much heavier than $\nu_{\al R}$. Hence, the lightest particle of $\nu_{\al R}$, such as $\nu_{1R}$ without loss of generality, and $\nu_{3R}$ play the role of two-component dark matter candidates. 

\begin{figure}[h]
\includegraphics[scale=1]{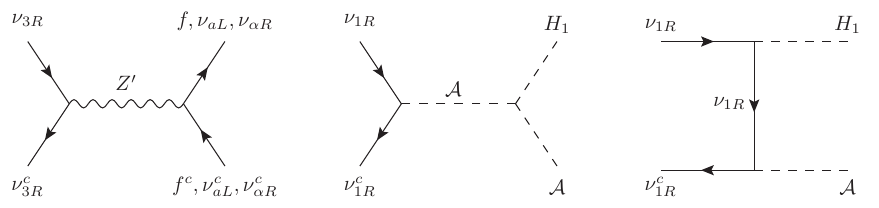}
 \caption[]{\label{fig5}Dominant contributions to annihilation of the two-component fermion dark matter.}
\end{figure}

The dominant channels in the dark matter pair annihilation of $\nu_{3R}$ ($\nu_{1R}$), if allowed by kinematics, are $\nu_{3R}\nu_{3R}^c \to ff^c,\nu_{aL}\nu_{aL}^c,\nu_{\al R}\nu_{\al R}^c,H_\al H_\al,\mathcal{A}\mathcal{A},\mathcal{S}_\al\mathcal{S}_\al,\mathcal{A}_\al\mathcal{A}_\al,Z'H_\al,Z'\mathcal{S}_\al,Z'\mathcal{A}_\al$, $Z'Z'$ ($\nu_{1R}\nu_{1R}^c \to ff^c,\nu_{aL}\nu_{aL}^c,H_1 H_1,\mathcal{A}\mathcal{A},H\mathcal{A},H_1 \mathcal{A}$) with $f$ indicates the SM charged fermions, which are proceeded via the $s$-channel interchange of new gauge boson $Z'$ and new scalar $H_2$ ($H_1,\mathcal{A}$), and via the $t$-channel of $\nu_{3R}$ ($\nu_{1R}, \ph_1^0,\ph_2^-$). However, this work assuming $\La_2\gg\La_1\gg v$, so only some annihilation channels of them are allowed. Additionally, for for the sake of simplicity, we assume that the relic density of $\nu_{3R}$ is governed mainly by $s$-channel exchange diagrams of $Z'$, while the pair annihilation processes of $\nu_{1R}$ are proceed dominantly though $s$- and $t$-channel exchange diagrams of $\mathcal{A}$ and $\nu_{1R}$ respectively, which are shown in Fig. \ref{fig5}. Here, there is the conversion between dark matter components, namely the heavier dark matter component $\nu_{3R}$ annihilates into the lighter one $\nu_{1R}$.

In the non-relativistic approximation, the thermal average annihilation cross-section times relative velocity for each dark matter component is estimated as
\bea \langle\sigma v\rangle_{\nu^c_{1R}\nu_{1R}\to H_1\mathcal{A}} &\simeq& \frac{(f_{11}^\nu)^2|4m^2_{\nu_{1R}}-m^2_{H_1}|}{512\pi m^4_{\nu_{1R}}} \left(|f_{11}^\nu|+\frac{\la_1\La_1}{2m_{\nu_{1R}}}\right)^2,\\
\langle\sigma v\rangle_{\nu^c_{3R}\nu_{3R}\to\mathrm{all}} &\simeq& \fr{(C^{Z'}_{\nu_{3R}})^2m^2_{\nu_{3R}}}{4\pi (4m^2_{\nu_{3R}}-m^2_{Z'})^2} \left[\sum_f N_C(f)(C^{Z'}_f)^2 +\fr{3(C^{Z'}_{\nu_L})^2}{2}+(C^{Z'}_{\nu_{\al R}})^2\right],
 \eea
where $f$ presents the SM charged fermions and the relevant couplings are shown in Eq. (\ref{couplings}). 

In the two-component dark matter framework, the dark matter relic abundance due to the thermal freeze-out is computed by solving the coupled Boltzmann equations. Assuming that the production of the $\nu_{1R}$ component from the $\nu_{3R}$ component is less significant compared to its annihilation into the $H_1,\mathcal{A}$ final states, one can obtain the individual relic abundance of each dark matter component through an approximate analytic solution as \cite{Bhattacharya:2016ysw}
\be \Omega_{\nu_{1R}}h^2 = \frac{1.07\times 10^9 x_{\nu_{1R}}}{\sqrt{g_\ast}m_\mathrm{P}\langle\sigma v\rangle_{\nu^c_{1R}\nu_{1R}\to H_1\mathcal{A}}},\hs \Omega_{\nu_{3R}}h^2 = \frac{1.07\times 10^9 x_{\nu_{3R}}}{\sqrt{g_\ast}m_\mathrm{P}\langle\sigma v\rangle_{\nu^c_{3R}\nu_{3R}\to\mathrm{all}}},\ee
where $g_\ast =106.75$ is the effective total number of degrees of freedom, $x_{\nu_{1,3R}}\simeq 25$ are parameters related to the freeze-out temperature. Hence, the dark matter relic abundance is sum of the individual contributions, namely
\be \Omega_{\mathrm{DM}}h^2=\Omega_{\nu_{1R}}h^2+\Omega_{\nu_{3R}}h^2. \ee

\begin{figure}[h]
\includegraphics[scale=0.42]{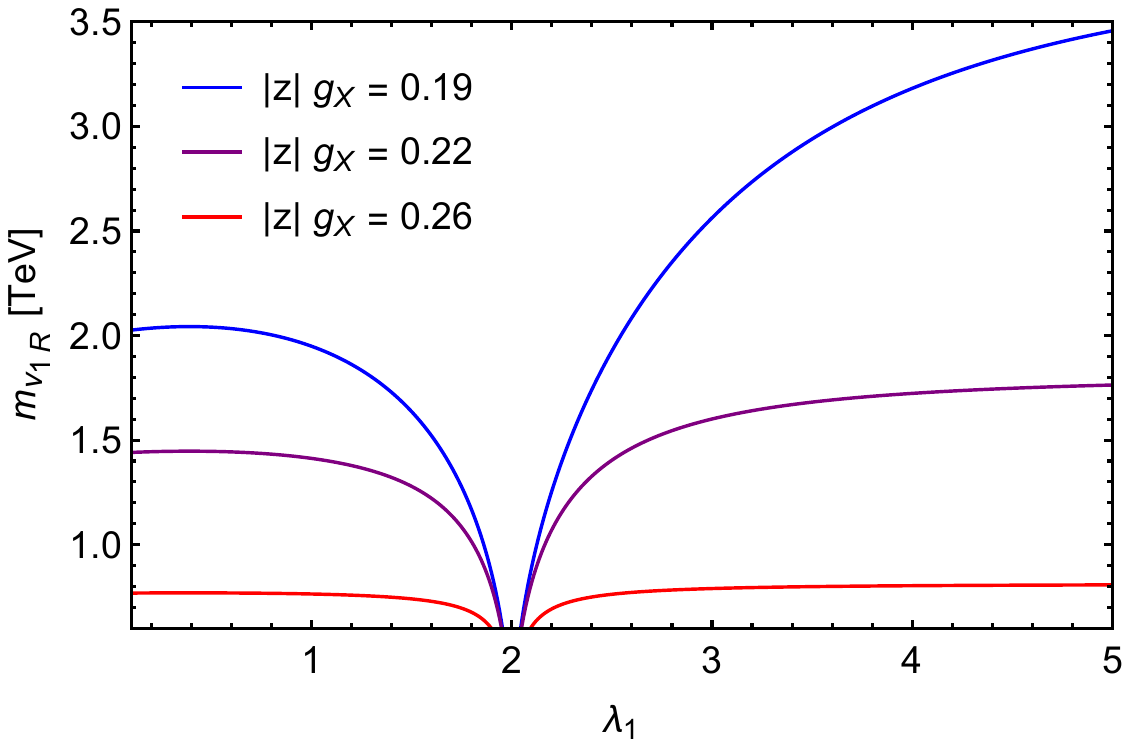}
\includegraphics[scale=0.42]{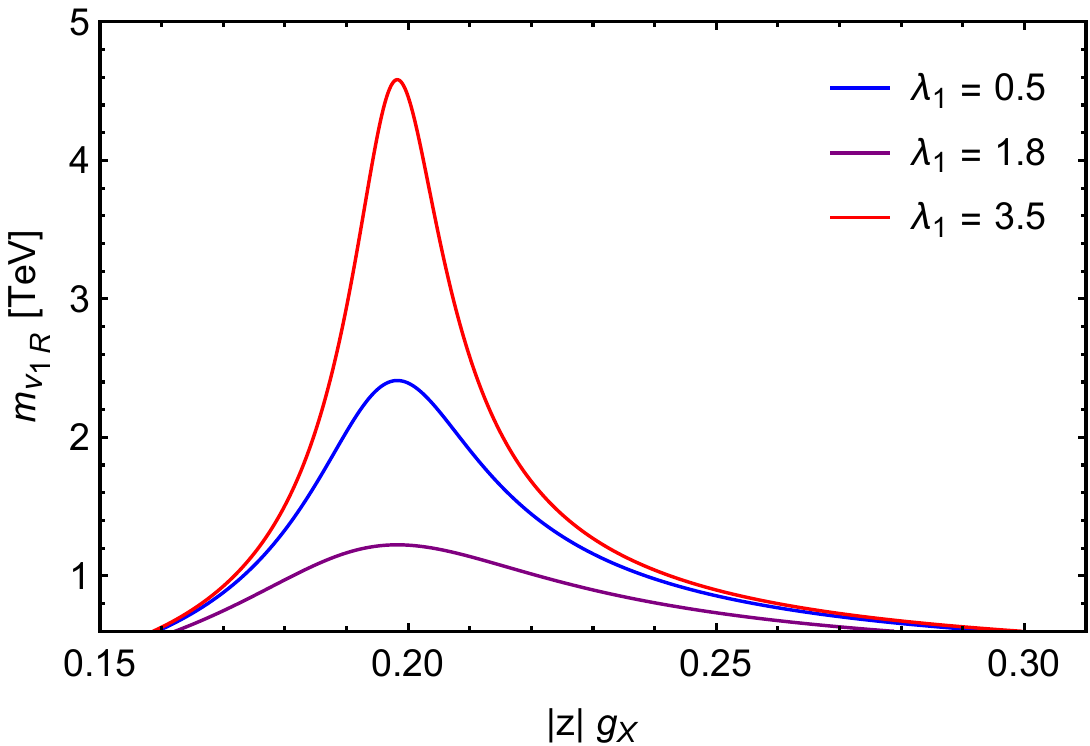}
 \caption[]{\label{fig6}Total dark matter density contoured as functions of dark matter masses and $\la_1$ (left panel) and $|z|g_X$ (right panel).}
\end{figure}

The relic density of dark matter depends on six parameters, which are $m_{\nu_{1R}}$, $m_{\nu_{3R}}$, $f^\nu_{11}$, $f^\nu_{33}$, $\la_1$, and the product $|z|g_X$. Taking $f^\nu_{11}= f^\nu_{33}=-\sqrt2$ and $m_{\nu_{3R}}=6 m_{\nu_{1R}}$, in Fig. \ref{fig6}, we make contours of the correct relic density of dark matter, i.e., $\Omega_{\mathrm{DM}}h^2\simeq 0.12$ \cite{Planck:2018vyg}, as a function of $m_{\nu_{1R}}$ and $\la_1$ for different choices of $|z|g_X$ (left panel), of $m_{\nu_{1R}}$ and $|z|g_X$ for different choices of $\la_1$ (right panel). This figure shows the viable mass region of $\nu_{1R}$ in the order $\mathcal{O}(1)$ TeV, and $0.16 \lesssim |z|g_X\lesssim 0.3$. Note that when $m_{\nu_{1R}}=m_{H_1}/2$, equivalently $\la_1=2$ (with $f^\nu_{11}=-\sqrt2$), then $\langle\sigma v\rangle_{\nu^c_{1R}\nu_{1R}\to H_1\mathcal{A}}=0$, which suppress the correct density and leads to a disconnected region on each curve in the left panel. In the right panel, when $m_{\nu_{3R}}=m_{Z'}/2$, equivalently $|z|g_X\simeq 0.2$ (with $\La_2\gg\La_1$), then $\langle\sigma v\rangle_{\nu^c_{3R}\nu_{3R}\to\mathrm{all}}\to \infty$, which reduces the relic density of $\nu_{3R}$ to zero, thus the correct relic density of dark matter is only set by the $\nu_{1R}$ pair annihilation. 

\begin{figure}[h]
\includegraphics[scale=0.5]{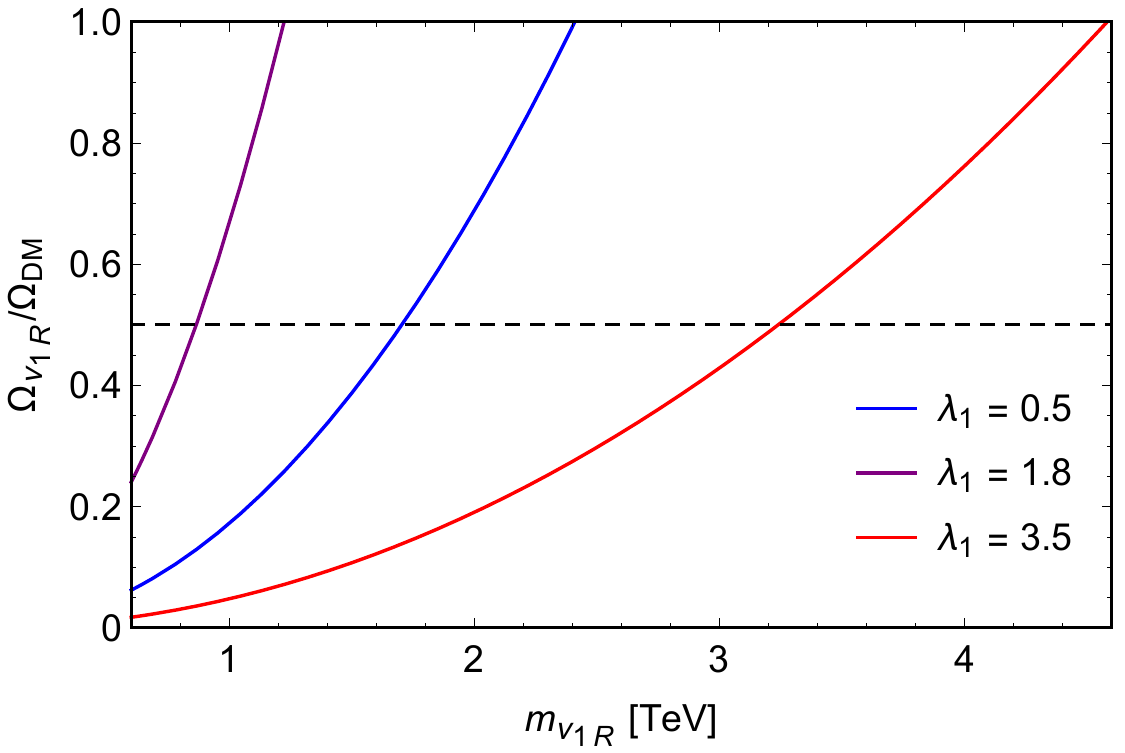}
 \caption[]{\label{fig7}The contribution ratio of dark matter components to the total density as function of dark matter masses.}
\end{figure}

In Fig. \ref{fig7}, we show the contribution ratio of dark matter components to the correct density as a function of $m_{\nu_{1R}}$ for different choices of $\la_1$, while $m_{\nu_{3R}}=6m_{\nu_{1R}}$ and $f^\nu_{11}=-\sqrt2$ kept fixed. The dashed black line corresponds to a contribution ratio of 0.5. It is easy to see that $\nu_{1R}$ dominantly contributes to the correct density when its mass takes on large values in the possible mass region.

Concerning the constraint from the direct detector experiments, it is interesting that two dark matter candidates under consideration are all Majorana fermions, which scatter with nucleon only via spin-dependent (SD) effective interactions by the $t$-channel exchange of the new gauge boson $Z'$. Such interactions give the SD scattering cross section of each dark matter candidate on a neutron target, $\sigma_{\mathrm{SD}}\sim g^{Z'}_A(\nu_{1,3R})[g^{Z'}_A(u)\la_u+g^{Z'}_A(d)(\la_d+\la_s)]$, where $g^{Z'}_A(f)$ is the axial-vector couplings of $Z'$ to $f$, while $\la_{u,d,s}$ are the fractional quark-spin coefficients \cite{Barger:2008qd}. Because the SM quarks are all vector-likes under $U(1)_X$, the axial-vector couplings of $Z'$ to them vanish. Hence, the model predicts a negative search result, as reported by the experiments \cite{XENON:2019rxp,PandaX-II:2018woa,LUX:2017ree}.

\section{\label{sec7}Conclusion}

In this work, we have proposed a $U(1)_X$ extension of the SM, where the $X$ charge depends on flavor and is a linear combination of the usual baryon and lepton numbers. By the anomaly cancellation conditions associated with the new charge $X$, this extension not only explains the existence of only three fermion families as observed but also requires the presence of at least three right-handed neutrinos similar to the SM extension with $U(1)_{B-L}$. 

Depending on the $X$ charge assignment for the right-handed neutrinos, we obtain two potential models for this approach, each of them predicts a distinct mechanism for neutrino mass generation, seesaw vs. scotogenic, as well as the dark matter structure, single- vs. two-component(s). Additionally, one of dark matter is stabilized by residual symmetry $p=(-1)^{X/z+2s}$, while the other one of dark matter is accidentally stabilized by the gauge symmetry. Both the models predict the dark matter nature as WIMP(s) with dark matter mass in the TeV regime. 

We have also investigated the FCNC and particle collider bounds and predict generically a lower bound for the new physics scale to be roundly $14.14$ TeV.

\section*{Acknowledgement}

DVL acknowledges the support by NAFOSTED under Grant No. NCNL.02-2022.61.   

\bibliographystyle{JHEP}

\bibliography{combine}

\end{document}